\documentclass{aa}
\usepackage{amsmath}
\usepackage{amssymb}
\usepackage{graphicx}
\usepackage{times}
\usepackage{natbib}

\usepackage{psfrag} 

\bibpunct{(}{)}{;}{a}{}{,} 

\begin{document}
%
%
\title{Hybrid Characteristics: 3D radiative transfer for parallel adaptive mesh refinement hydrodynamics}
\titlerunning{Hybrid Characteristics: 3D radiative transfer for parallel AMR hydrodynamics}
\author{E.-J. Rijkhorst\inst{1} \and T. Plewa\inst{2,3} \and A. Dubey\inst{2,3} \and G. Mellema\inst{4,1}}
\institute{Sterrewacht Leiden, P.O. Box 9513, 2300 RA, Leiden, The Netherlands
\and Center for Astrophysical Thermonuclear Flashes, University of Chicago, 5640 South Ellis Avenue, Chicago, IL 60637
\and Department of Astronomy \& Astrophysics, University of Chicago, 5640 South Ellis Avenue, Chicago, IL 60637
\and ASTRON, P.O. Box 2, 7990 AA Dwingeloo, The Netherlands}
\offprints{E.-J. Rijkhorst\\\email{rijkhorst@strw.leidenuniv.nl}}
\date{Received * / Accepted *}
%
%
\abstract{
We have developed a three-dimensional radiative transfer method designed specifically for use with parallel adaptive mesh refinement hydrodynamics codes.
This new algorithm, which we call hybrid characteristics, introduces a novel form of ray tracing that can neither be classified as long, nor as short characteristics, but which applies the underlying principles, i.e. efficient execution through interpolation and parallelizability, of both.

Primary applications of the hybrid characteristics method are radiation hydrodynamics problems that take into account the effects of photoionization and heating due to point sources of radiation.
The method is implemented in the hydrodynamics package FLASH.
The ionization, heating, and cooling processes are modelled using the DORIC ionization package.
Upon comparison with the long characteristics method, we find that our method calculates the column density with a similarly high accuracy and produces sharp and well defined shadows.
We show the quality of the new algorithm in an application to the photoevaporation of multiple over-dense clumps.

We present several test problems demonstrating the feasibility of our method for performing high resolution three-dimensional radiation hydrodynamics calculations that span a large range of scales.
Initial performance tests show that the ray tracing part of our method takes less time to execute than other parts of the calculation (e.g. hydrodynamics and adaptive mesh refinement), and that a high degree of efficiency is obtained in parallel execution.
Although the hybrid characteristics method is developed for problems involving photoionization due to point sources, the algorithm can be easily adapted to the case of more general radiation fields.
%
%
}
%
%
\maketitle
%
%
\section{Introduction}
Current multi-dimensional parallel adaptive mesh refinement (AMR, see \citet{Berger1984,Berger1989}) hydrodynamics codes, include more and more physical processes like \mbox{(self-)} gravity, nuclear burning, and composition dependent equations of state.
Furthermore, a wealth of different solvers for relativistic or magneto-hydrodynamics, have become available.
These codes are in general implemented as a modular framework, facilitating a rather straightforward inclusion of new physics modules, and are often distributed freely for scientific use \citep{2000ApJS..131..273F, 2004OSheaAMRConf, 2000RMxAC...9...66N}.

Since astrophysical applications are many times dominated by radiative processes, it is highly desirable that radiative transfer in some form is included in these codes.
Efforts to solve the full equations of radiative transfer \citep[using the Eddington tensor formalism in combination with short characteristics, see][]{1992ApJS...80..819S}, or in the flux-limited diffusion approximation \citep{2001ApJS..135...95T,2004MNRAS.353.1078W}, together with the hydrodynamics have been made, but it remains a complex task to create a parallel algorithm which combines radiative transfer and hydrodynamics for multi-dimensional calculations that runs efficiently on todays multi-processor supercomputers \citep[e.g.][]{2003ApJS..147..197H}.

For many astrophysical applications however, it is not necessary to solve the full set of radiative transfer equations; for these specific cases it is sufficient to just determine the optical depth due to absorption along a line of sight from the source to a certain location in the computational domain.
For the purpose of our application of ionization calculations, the optical depth is used to determine the photoionization and heating rates.
When this is combined with detailed calculations of radiative cooling, many applications come within reach, such as the evolution of planetary nebulae \citep{1994A&A...289..937F}, photoevaporation of cosmological mini-haloes \citep{2004MNRAS.348..753S}, photoevaporation of cometary knots \citep{2003A&A...405..189L}, the evolution of proplyds \citep[e.g.][]{2000ApJ...539..258R}, or even simplified scenarios of explosions of massive stars \citep{1996A&A...306..167J}, to name just a few.

In creating a method that combines radiative transfer and hydrodynamics, one in general starts with an existing hydrodynamics code and adds the necessary radiation processes to it \citep[e.g.][]{1998A&A...331..335M,2001ApJS..135...95T,2004MNRAS.353.1078W,2005AstrophHeinemann,2005ApJ...620..840L}.
In this paper we describe the addition of a new radiative transfer algorithm, which we call {\em hybrid characteristics}, to the parallel 3D AMR hydrodynamics package FLASH \citep{2000ApJS..131..273F}.

Most of the radiative transfer methods that were successfully combined with extant hydrodynamics codes apply some form of ray tracing to find the optical depth at each location in the computational domain.
Apart from ray tracing one could also use statistical methods to find the solution to the radiative transfer equations \citep[e.g.][]{2003MNRAS.345..379M}.
Yet another approach could be the use of Fourier transforms \citep{2002ApJS..141..211C}, or unstructured grids \citep{2004AstrophRitzerveld}, to determine the radiation field.

For a one-dimensional, non-AMR, serial code, ray tracing becomes a rather straightforward procedure which requires little second thought.
Equivalently, the case of a plane parallel radiation field on a Cartesian grid, or a single point-source at the centre of a spherically symmetric grid, for which all rays run parallel to a coordinate axis, can be handled quite easily.
Although this type of implementation can readily be used to study a number of interesting astrophysical phenomena, it is still highly desirable to have a code that can treat the more general case of a point source of ionizing radiation on a 3D Cartesian domain.
Such more general methods were for example implemented by \citet{2000MNRAS.314..681R,2000ApJ...539..258R,2003A&A...405..189L}, but none of these methods was explicitly parallelized for distributed memory machines though.

The aim of this work is to create a characteristics-based radiative transfer method that can handle multiple sources of ionizing radiation in AMR enabled simulations to be run on distributed memory parallel machines.
For this, a radical rethink of the concept of ray tracing is necessary, since, for this type of parallel AMR codes, the computational domain is not only sub-divided into a hierarchy of patches, but is also distributed over a number of processors.
The first choice one therefore has to make is which flavour of ray tracing one wants to apply: either long or short characteristics.
Since these two methods have rather different properties when it comes to efficiency and parallelizability, this choice will determine the success of the final algorithm.

We are aware of a number of other methods that use some form of adaptivity to solve the radiative transfer equations:
\citet{2002MNRAS.330L..53A} designed a method where the ray itself is adaptively split into sub-rays, but the underlying grid is still regular.
\citet{2002JQSRT..75..765S} employed second order finite differencing of the full radiative transfer equations on an oct-tree AMR grid, and, more recently, \citet{2005ApJ...618..744J} implemented a ray tracing method for cell-based AMR.
\citet{Jessee1998} presented a radiative transfer method for patch-based AMR that uses the discrete ordinates approach.
However, none of these methods resulted in parallel algorithms used in applications in which radiation is coupled to hydrodynamics.

Efforts to create a parallel radiation hydrodynamics code were presented by e.g. \citet{2001MNRAS.321..593N,2003ApJS..147..197H}, and, more recently, a three-dimensional method by \citet{2005AstrophHeinemann}, who developed a ray tracing algorithm for decomposed domains.
However, none of these two methods uses AMR.

Our presentation begins with Sect. \ref{sect:charBasedAlgo} in which we describe our new method.
This method can {\em not} be classified as either short or long characteristics, but does have the desired properties, namely high parallel and computational efficiency, of its two predecessors.
We also compare our method to two recent ones which share similar features with ours.
Supplemental physics components required by our primary target application (gas ionization, heating, and cooling) are presented in Sect. \ref{sect:ionizeHeatCool}, where we give a brief description of the DORIC routines \citep[][and references therein]{2002A&A...394..901M}.
In Sect. \ref{sect:testing} we first compare the accuracy with which our method calculates column densities to results obtained with a standard long characteristics approach.
Then we present a pure radiation transport problem aimed at testing the accuracy of the ionization state calculations and shadow casting.
This is followed by a coupled radiation hydrodynamics calculation of a photoevaporation flow.
Sect. \ref{sect:performance} presents some initial performance results.
Discussion of possible extensions and future applications for our method together with the conclusions are given in Sect. \ref{sect:conclusions}.
%
%
\section{Characteristics based radiative transfer}
\label{sect:charBasedAlgo}
%
%
\begin{figure}
\psfrag{a}{\bf a}
\psfrag{b}{\bf b}
\resizebox{\hsize}{!}{\includegraphics[width=10cm]{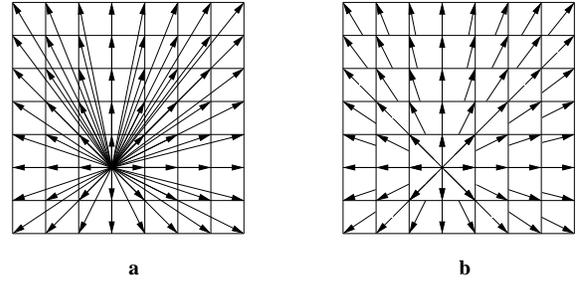}}
  \caption{
Comparing the long (a) and short (b) characteristics method.
For the long characteristics method, the closer one gets to the source, the more rays pass through (approximately) the same part of a cell, resulting in a large number of redundant calculations.
The short characteristics method does not suffer from this, since here column densities are interpolated from cells that have been dealt with previously, so only the contributions to the column density of the short ray sections that pass from cell to cell need to be computed.
}
  \label{fig:longShortCompare}
\end{figure}
%
%
When calculating the effects of ionizing radiation due to a point source, the radiation field is often dominated by this source, and one can safely ignore contributions to the radiation field due to the ambient gas.
This means that the radiative transfer equations assume a particularly simple form, since we can take the total emission coefficient (and thereby the source function) to be equal to zero.
Furthermore, when we also ignore the effects of scattering, the solution to the radiative transfer equations for the specific intensity $I$ at location $\boldsymbol{r}$ is given by
\begin{equation}
  I(\boldsymbol{r}) = I(0) \exp(-\tau(\boldsymbol{r})) \: , 
\end{equation}
and only depends on the optical depth $\tau$, which is defined by
\begin{equation}
  \tau(\boldsymbol{r}) = a_0 N(\boldsymbol{r}) \: ,
\label{eq:opticaldepth}
\end{equation}
with $a_0$ the absorption cross section, and $N$ the column density at $\boldsymbol{r}$.

Once the optical depth is known at every location in the computational domain, one can use it to find the ionization, heating, and cooling rates, and calculate the ionization state and temperature of the gas.
Since, for finite-volume hydrodynamics codes, the computational domain is discretized into cells, the optical depth, or, equivalently, the column density for a certain cell, is found by adding the contributions from all cells that lie between the source and the destination cell under consideration.
This can be achieved by casting a ray, or {\em long characteristic}, from the source to the cell, accumulating contributions to the total column density along the way.
In case of an AMR hierarchy, the algorithm first needs to identify the patches and cells contained within the patches that are traversed by the ray, and then calculate their local contributions to the total column density.

Although the method of long characteristics is very accurate, it is also rather inefficient, since, the closer a cell is to the source, the more rays cut through (approximately) the same part of the cell, introducing a lot of redundant calculations (see Fig. \ref{fig:longShortCompare}a).
A way to eliminate this redundancy is to use the method of {\em short characteristics}.
Here, the total column density for a certain cell is calculated by interpolating upwind values of column density calculated in a previous step, thereby creating some diffusion, but removing the redundant calculations inherent in the long characteristics method (Fig. \ref{fig:longShortCompare}b).
For this to work, the appropriate information from upwind cells needs to be available at all times, which means one needs to sweep the numerical grid outwards from the source.
This necessity of having to traverse the grid in a certain order makes this method intrinsically serial, since values of column density in cells now depend on one another.
The long characteristics method does not suffer from this restriction, because here contributions to the total column density from cells cut by a ray do not depend on column densities in other cells.
Therefore, the long characteristics method is fully parallelizable, since calculations of contributions to the column density along each ray can be performed independently.
For our method we combine the desirable qualities of both these approaches; the idea of interpolation is adopted from the short characteristics method, while parallelism is obtained following principles of the long characteristics method.

In what follows, we start with a general description of the algorithm used to trace rays on AMR hierarchies.
We explain how the long characteristics method is exploited to make this a parallel algorithm, and where the interpolation comes in to increase the efficiency of the calculation.

Although our algorithm is designed for three dimensions, many features of its implementation can be explained using two-dimensional analogues.
Whenever the generalization from two to three dimension is non-trivial, we will supply the full, three-dimensional, description.
Since the algorithm is naturally subdivided into a number of steps, we will expand on these separately.
%
%
\subsection{The distributed computational domain}
\begin{figure}
\psfrag{processor 0}{\bf processor 0}
\psfrag{processor 1}{\bf processor 1}
\resizebox{\hsize}{!}{\includegraphics[width=10cm]{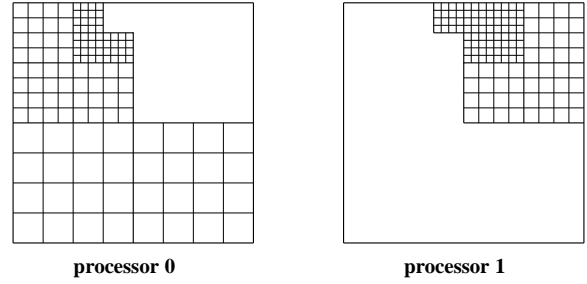}}
  \caption{
Two-dimensional example of an AMR hierarchy distributed over two different processors.
Here, each patch contains $4\times 4$ cells.
}
  \label{fig:totDomain}
\end{figure}

Consider a computational domain that is distributed over $N_p$ processors (for a two-dimensional example, see Fig. \ref{fig:totDomain}).
%
%
%
%
Rays are traced over these different sub-domains and must therefore be split up into independent {\em ray sections}.
Naturally, these sections are in the first place defined by the boundaries of each processor's sub-domain, and in the second place by the boundaries of the patches contained within that sub-domain.

So first each processor calculates for all the patches it owns the {\em local} column densities $\Delta N$.
These local contributions are found by tracing ray sections that originate at the patch faces that are located {\em closest} to the source, and that terminate at the centres of the cells (see Fig. \ref{fig:raySections}).
Since finding these contributions is a local process, this part of the algorithm is fully parallel, and can be implemented using either the short or long characteristics method.
Details on how the ray tracing for individual patches is implemented are given in Sect. \ref{sect:rayTraceLocal}.
Note however that, before each processor can calculate its $\Delta N$, it needs to know the physical location of the source, so this information is made available first.

Since in general rays traverse more than one processor domain, exchange of information has to take place at some point in the algorithm.
After this communication step has finished, each processor should have available all contributions of column density to the rays that terminate in its domain.
By interpolating and accumulating all these contributions for all rays, one ultimately obtains the total column density for each cell (see Sect. \ref{sect:communicate} for a more elaborate description of the communication patterns involved).
Details on the procedure applied to find the patches cut by a ray, and the way in which their contributions to the total column density are subsequently calculated, are given in Sect. \ref{sect:cutList} and Sect. \ref{sect:accumulate}, respectively.
%
%
\subsection{Ray tracing a single patch}
\label{sect:rayTraceLocal}
%
%
\begin{figure}
\psfrag{a}{\bf a}
\psfrag{b}{\bf b}
\resizebox{\hsize}{!}{\includegraphics[width=10cm]{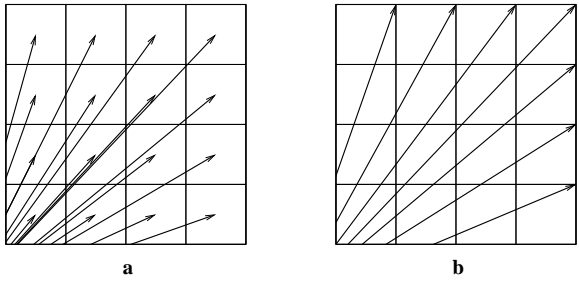}}
  \caption{
Two-dimensional example of ray sections for a single patch.
Local contributions to the column density are indicated by ray sections that terminate at cell centres (a), whereas contributions that are to be communicated between processors, and are subsequently used in an interpolation step, terminate at cell corners (b).
The source lies outside of the patch in the direction of the lower left corner.
}
  \label{fig:raySections}
\end{figure}
%
%
In this section we will explain how the contribution to the total column density along a local ray section in a single patch can be calculated (see Fig. \ref{fig:raySections} and Fig. \ref{fig:summary}).
As explained above, each patch can be dealt with independently, which makes this part of the calculation fully parallel.

The local column density contributions are calculated from
\begin{equation}
  \Delta N = \sum_\mathrm{cells} x(\mathrm{HI}) n(\mathrm{H}) \Delta s \: ,
\end{equation}
with $x(\mathrm{HI})$ the ionization fraction of neutral hydrogen, $n(\mathrm{H})$ the hydrogen number density, and $\Delta s$ the physical path length through the cell.

These contributions are found by casting a ray section from the faces of the patch that are located closest to the source towards each cell centre (see Fig. \ref{fig:raySections}).
Column density contributions by the cells that lie inside the patch along each section are calculated using the `fast voxel traversal algorithm' from \citet{1987Eurographics..1A} (for more details on this traversal method, see App. \ref{app:voxels}).
Besides ray sections that terminate at cell centres, we also need to calculate the column density contribution for ray sections that lead to cell {\em corners} located at those patch faces that are farthest away from the source (see Fig. \ref{fig:raySections}).
These are the contributions to the column density that need to be communicated (Sect. \ref{sect:communicate}), and interpolated (Sect. \ref{sect:accumulate}) in subsequent steps of the algorithm.

Calculating these ray sections is similar to the method of long characteristics, but since the number of cells per patch is low relative to the effective resolution of the full computational domain, this actually does not impair the performance of the method too much (see Sect. \ref{sect:performance} for an analytical comparison of our method with the short and long characteristics one for a regular grid).

We also considered using short instead of long characteristics to ray trace a single patch (see App. \ref{app:shortChar} for a description of a possible implementation).
However, although the short characteristics method executes presumably more efficiently than the long characteristics one, the first requires interpolation, whereas the latter simply adds up column density contributions by individual cells.
When the number of cells that need to be traversed is relatively small, as is the case when ray tracing the single patches, these extra calculations may render the short characteristics method even less efficient than the long characteristics one.
Furthermore, the interpolation introduces undesirable diffusion.
We therefore decided to implement the more accurate and straightforward ray tracing approach of \citet{1987Eurographics..1A}. 
%
%
\subsection{Hybrid Characteristics}
\label{sect:hybridChar}
As was mentioned above, in AMR hydrodynamics codes, each processor owns a sub-domain of the computational volume which is covered by a collection of patches.
In order to obtain the total column density for a certain ray that traverses these sub-domains, individual local contributions by the patches need to be accumulated.
This can be interpreted as applying the method of long characteristics, in this case {\em not} to add up contributions from individual cells, but instead to add up contributions from individual patches.
So here our algorithm does again make use of long characteristics but now at the level of patches.

Since each processor knows the direction of its rays and the co-ordinates where they terminate, it can find the patches cut by these rays and perform the required calculations.
For certain flavours of AMR, patches from different refinement levels may partially overlap.
In such cases, one would have to make sure that only parts of the patches that contain valid data (i.e., the data from regions resolved to the highest resolution) are considered in the calculation of the column density.
One way to eliminate the overlap is to apply a procedure called `grid homogenization', as described by \citet{KreylosWeberBethelShalfHamannJoy:2002-257}.

For the oct-tree type of AMR implemented in FLASH, patches from different refinement levels do not overlap.
Patches are either fully covered by still more refined patches or otherwise contain valid data (the latter are the so-called `leaf patches' in terminology of FLASH).
Therefore, a simple check to see if a patch is a `leaf patch' is sufficient to determine whether or not it should contribute to the total column density along the ray.

Once the list of patches traversed by a ray is known, we loop through it, and determine the local column density contributed by each patch to the total column density for the ray.
Unless the ray terminates in the patch under consideration, it will in general not exit a patch exactly at a cell corner.
This means that we need to interpolate the values of column density contribution $\Delta N$, obtained earlier (using either the short or long characteristics method as described in Sect. \ref{sect:rayTraceLocal}) at that face of the patch where the ray leaves it.

We would like to emphasize that, although our method makes use of some form of long characteristics, {\em nowhere} in the algorithm is a ray traced on a cell-by-cell basis over the {\em full} computational domain.
To the contrary, ray sections are traced through the cells of each patch and it is these local contributions which are combined through interpolation by performing another ray trace, this time not over cells but over patches, as described in Sect. \ref{sect:cutList}.
This is why we call our algorithm {\em hybrid characteristics}.

Below, we first explain how the local column density contributions $\Delta N$, obtained with one of the methods from Sect. \ref{sect:rayTraceLocal}, are communicated between processors.
Then we describe how the list of patches traversed by a ray is constructed, after which we show the way in which this list is used to calculate the contributions to the total column density $N$.
%
%
\subsubsection{Communicating local column density contributions}
\label{sect:communicate}
Since, for a parallel AMR hydrodynamics code, the patches are distributed over a number of processors, communication between processors is inevitable at certain points in the algorithm.
In particular, as soon as the local contributions to the column density have been calculated (Sect. \ref{sect:rayTraceLocal}), values of these $\Delta N$ located at patch faces that are farthest away from the source are communicated between processors.
In this way, each processor has the information regarding the face values of local column density from {\em all} patches in existence (i.e. the so called `gather' operation is used).
Apart from these face values, all processors also need information about the location and size of each patch and its refinement level in order to determine if a particular ray cuts a patch.
This information is communicated using the `gather' operation as well.

The size of the messages to be communicated and the memory needed for storage of this information is given by
\begin{equation}
  P_\mathrm{tot} \, p_\mathrm{max} \, S \: ,
\end{equation}
where $P_\mathrm{tot}$ is the total number of processors, $p_\mathrm{max}$ is the maximum number of patches in existence on any processor, and $S$ is the required storage space per patch.
In three dimensions, $S$ should contain the values of $\Delta N$ from the three patch faces located farthest away from the source, as well as the location, size, and refinement level information of each patch.

For an initial test of the performance of the algorithm as a whole, and of its communication patterns in particular, see Sect. \ref{sect:performance}.
%
%
\subsubsection{Constructing the list of patches cut by a ray}
\label{sect:cutList}
A straightforward approach to constructing the list of patches traversed by a ray would be to simply check for all patches whether or not they are cut by the ray under consideration.
Since this would have to be done for all rays, and since there are as many rays as there are cells, this approach quickly becomes prohibitively slow.
We therefore developed a new, more elaborate, but much faster method to find the list of patches cut by a ray.

\begin{figure}
\psfrag{processor 0}{\bf processor 0}
\psfrag{processor 1}{\bf processor 1}
\psfrag{1}{1}
\psfrag{2}{2}
\psfrag{3}{3}
\psfrag{4}{4}
\psfrag{5}{5}
\psfrag{6}{6}
\psfrag{7}{7}
\psfrag{8}{8}
\psfrag{9}{9}
\psfrag{10}{10}
\psfrag{11}{11}
\psfrag{12}{12}
\psfrag{13}{13}
\psfrag{14}{14}
\psfrag{15}{15}
\psfrag{16}{16}
\resizebox{\hsize}{!}{\includegraphics[width=10cm]{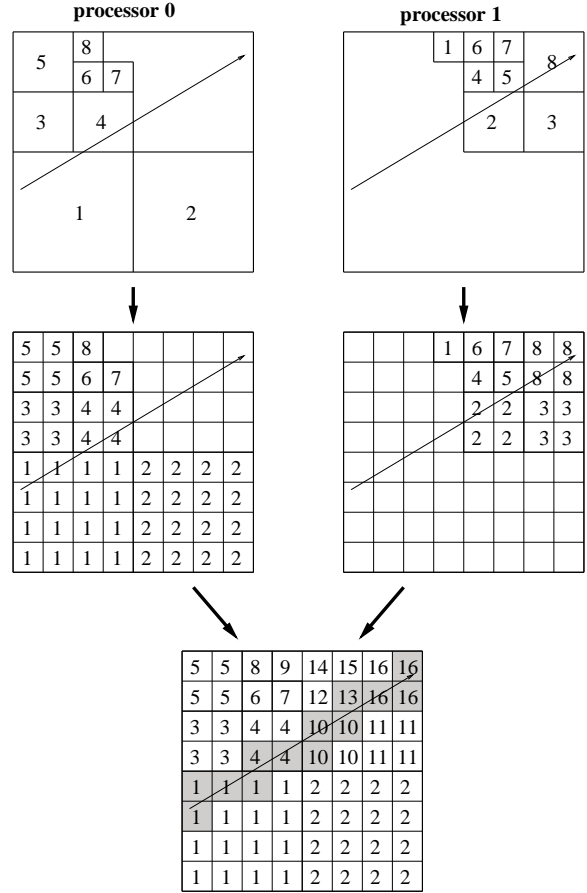}}
  \caption{
Two-dimensional example of a `patch-mapping' for a computational domain that is split over two different processors.
In the top row the local ids of the patches on the different processors are shown.
The mapping of these patch ids onto the patch-mapping array is shown in the middle row.
The bottom row shows the global patch-mapping after the local patch-mappings have been communicated.
Tracing the depicted ray results in the patch list $\{1,4,10,13,16\}$.
The patch-mapping entries visited during the ray tracing are shown in grey.
}
  \label{fig:gridMapping}
\end{figure}

First, each processor creates a so called `patch-mapping' which consists of an integer array representing the full computational domain that stores the id (i.e. a unique integer identifier) of all patches containing valid data.
In Fig. \ref{fig:gridMapping} we show an example of such a mapping.
%
%
%
%
These local patch-mapping arrays then need to be communicated and merged (using a so called `reduce' communication operation) after which each processor has the same global patch-mapping corresponding to the full computational domain.

In order to discern patches that are on different processors we use the following coding for the global patch id:
\begin{equation}
  p_G = p_L + P \, p_\mathrm{max} \: ,
\end{equation}
with $p_G$ the global patch id, $p_L$ the local patch id, and $P$ the processor id.

We then trace the ray, again using the `fast voxel traversal algorithm' (\citet{1987Eurographics..1A}, see App. \ref{app:voxels}), but now to trace through the {\em global patch-mapping array}.
This results in the list of patches cut by the ray, which is used to accumulate their local contributions, which were already communicated earlier, to arrive at the total column density (as described in Sect. \ref{sect:accumulate}).

Although this approach to ray tracing can be a potential bottle-neck in the algorithm, one needs to keep in mind that the maximum number of patch-mapping entries along a ray is given by $\sqrt{3}C/c$, with $C^3$ the total number of cells if the computational domain would be fully refined, and $c^3$ the number of cells per patch.

For a typical three-dimensional oct-tree type AMR simulation with $C=512$ and $c=16$, we find a maximum amount of $\sim 55$ patch-mapping entries that are cut by a ray.
Note however that this is an upper limit.
The number of entries is drastically smaller when the source and destination of the ray are not located at opposite sides of the domain (which will be the case for most rays).
Note also that, although we have to trace through the patch-mapping entries, the actual number of patches that ends up in the list is strongly reduced due to the adaptive nature of the discretization.
In the example given in Fig. \ref{fig:gridMapping}, the number of patch-mapping entries visited by the ray is 13, but the number of patches that end up in the list is only 5.
It is this latter number which determines how many interpolations are needed when accumulating the local column density contributions.
%
%
\subsubsection{Accumulating local column density contributions}
\label{sect:accumulate}
Now that we have the list of patches traversed by a ray (Sect. \ref{sect:cutList}) and the values of local column density at the patch faces located farthest away from the source have been made available to all processors (Sect. \ref{sect:communicate}), we can proceed and calculate the local contributions to the total column density through interpolation.

The calculations that need to be performed for a ray $r$ traversing a patch $p$ can be broken up into the following steps (two-dimensional case, see Fig. \ref{fig:accumulate2D}):
%
%
\begin{figure}
\psfrag{l1}{$l_1$}
\psfrag{l2}{$l_2$}
\psfrag{le}{$l_e$}
\psfrag{c1}{$c_1$}
\psfrag{c2}{$c_2$}
\psfrag{e}{$e$}
\resizebox{\hsize}{!}{\includegraphics[width=10cm]{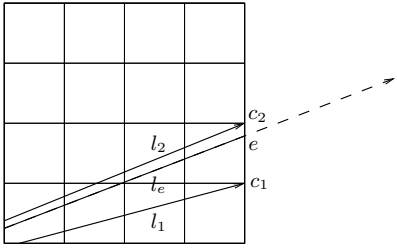}}
  \caption{
Two-dimensional illustration of the linear interpolation scheme used to accumulate local column density contributions.
Shown are the ray sections used in the interpolation (see text for further details).
}
  \label{fig:accumulate2D}
\end{figure}
%
%
%
%
\begin{figure}
\psfrag{e}{$e$}
\psfrag{1}{$1$}
\psfrag{2}{$2$}
\psfrag{3}{$3$}
\psfrag{4}{$4$}
\psfrag{5}{$5$}
\psfrag{6}{$6$}
\psfrag{7}{$7$}
\psfrag{8}{$8$}
\resizebox{\hsize}{!}{\includegraphics*[width=10cm]{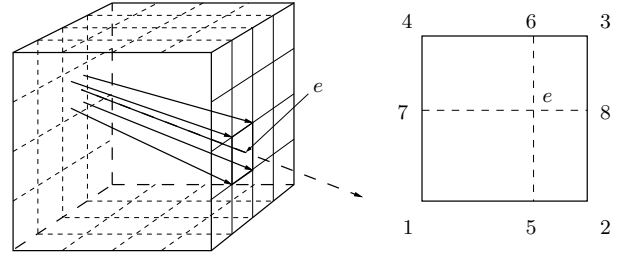}}
  \caption{
Illustration of the interpolation scheme in three dimensions.
For clarity we show outlines of cells on patch faces only.
In the left image we show a ray $r$ that exits the patch at location $e$ through a cell face, together with the ray sections used in the interpolation that terminate at the corners of this cell face.
The image on the right shows the cell face in more detail, where we indicated the cell corners by $1$, $2$, $3$, and $4$.
In addition to these cell corners, ray sections used in the interpolation that terminate at $5$, $6$, $7$, and $8$ are also indicated (see text for further details).
}
  \label{fig:accumulate3D}
\end{figure}
%
%
\newcounter{Lcount1}
\begin{list}{\arabic{Lcount1}}
{\usecounter{Lcount1}}
\item Find the location $e$ where $r$ exits $p$.
\item Use this to find the two cell corners $c_1$ and $c_2$ that are closest to $e$ and store their corresponding local column density contributions $\Delta N_1$ and $\Delta N_2$.
\item Calculate the geometrical path lengths of the ray sections that terminate in $c_1$, $c_2$, and $e$, and denote these by $l_1$, $l_2$, and $l_e$, respectively.
\item Use these path lengths to calculate the following normalized interpolation weights:
\begin{equation}
  w_1 = |l_2 - l_e|/(l_1 + l_2),\:
  w_2 = |l_1 - l_e|/(l_1 + l_2)\: .
\end{equation}
\item Calculate the desired value of local column density at $e$ through linear interpolation:
\begin{equation}
\Delta N_e = w_1 \, \Delta N_1 + w_2 \, \Delta N_2 \: .
\end{equation}
\end{list}
After all $\Delta N_e$ for each patch in the list of patches cut by $r$ are calculated, we simply need to sum them to arrive at the total column density for $r$:
\begin{equation}
\label{eq:totcoldens}
  N(r) = \sum_p \Delta N_e(p) \:\:\: [p\in \mathrm{list}(r)] \: .
\end{equation} 

The interpolation weights given above were constructed using the conditions
\begin{equation}
  w_1\, l_1 + w_2\, l_2 = l_e, \;\; \mathrm{and} \;\; w_1+w_2=1 \: ,
\end{equation}
which, for the case of a homogeneous density distribution, results in the {\em exact} solution for the column density (i.e., apart from a constant factor, the path length itself).
Other weights, like ones derived from the distances between the exit locations $e$, $c_1$, and $c_2$, can also be used, but this leads to $\sim 10\%$ errors for rays that enter a patch close to a patch corner (as is depicted by the example ray section of Fig. \ref{fig:accumulate2D}).

In three dimensions (see Fig. \ref{fig:accumulate3D}) it is not straightforward to derive weights that are a generalization of the two-dimensional ones described above.
We therefore give a more intuitive derivation of these weights, using a procedure where we apply the weights for the two-dimensional case twice in succession:
%
%
\begin{figure*}
\centering
\includegraphics*[width=17cm]{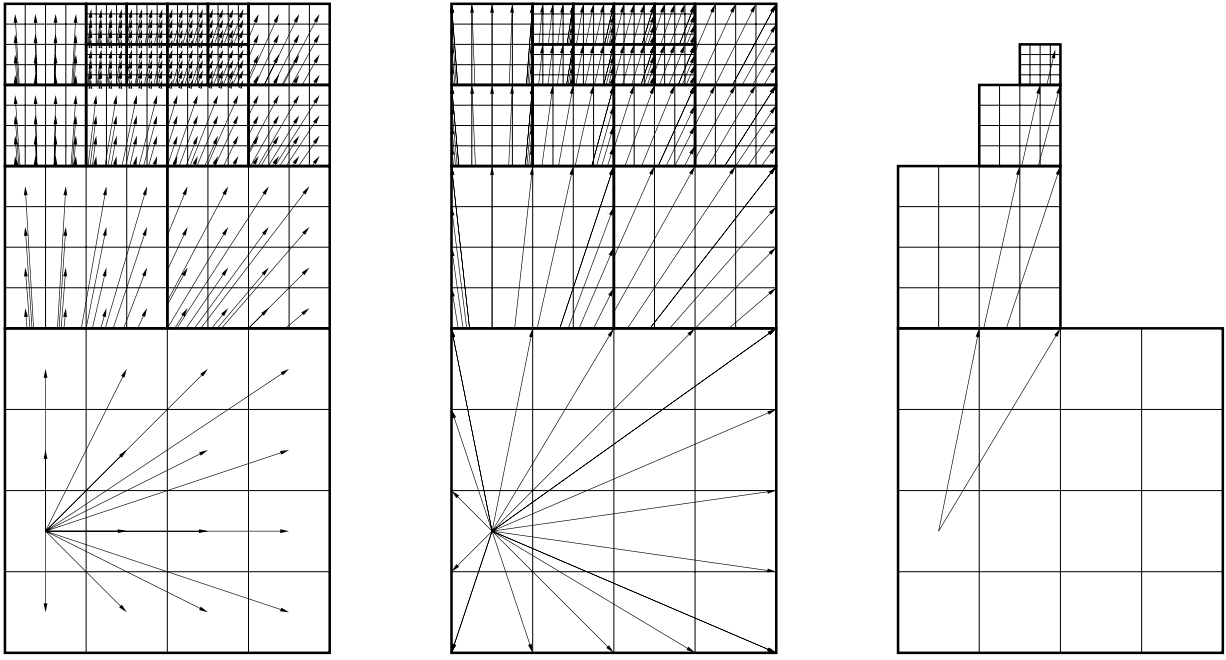}
  \caption{
Summary of steps taken in the hybrid characteristics method.
On the left we show ray sections that represent local contributions to the column density (summary step 3), whereas ray sections that represent values of column density that need to be communicated are shown at the centre image (summary step 4).
Note that only those values on patch faces located farthest away from the source need to be communicated.
On the right we show an example of the interpolation of these local values for a particular destination cell (summary step 6).
Note that there is no need to interpolate the value for the final ray section in the destination patch since its value was already calculated previously (summary step 3).
}
  \label{fig:summary}
\end{figure*}
%
%
\newcounter{Lcount2}
\begin{list}{\arabic{Lcount2}}
{\usecounter{Lcount2}}
\item Find the location $e$ where $r$ exits $p$.
\item Use this to find the four cell corners $c_1$, $c_2$, $c_3$, and $c_4$ that are closest to $e$ and store their corresponding local column density contributions $\Delta N_1$, $\Delta N_2$, $\Delta N_3$, and $\Delta N_4$.
\item Calculate the geometrical path lengths of the ray sections that terminate in $c_1$, $c_2$, $c_3$, $c_4$, and $e$ and denote these by $l_1$, $l_2$, $l_3$, $l_4$, and $l_e$, respectively.
Also calculate the path lengths $l_5$ and $l_6$ of the ray sections that terminate in $c_5$ and $c_6$ respectively (see Fig. \ref{fig:accumulate3D}).
\item Use these path lengths to calculate the following normalized interpolation weights:
\begin{equation}
\begin{array}{cc}
w_1 = |l_2 - l_5|/(l_1 + l_2), & w_2 = |l_1 - l_5|/(l_1 + l_2) \: , \\
w_3 = |l_4 - l_6|/(l_3 + l_4), & w_4 = |l_3 - l_6|/(l_3 + l_4) \; , \\
w_5 = |l_6 - l_e|/(l_5 + l_6), & w_6 = |l_5 - l_e|/(l_5 + l_6) \: .
\end{array}
\end{equation}
\item Calculate the values of local column density $\Delta N_5$ and $\Delta N_6$ at $c_5$ and $c_6$ through linear interpolation:
\begin{equation}
\begin{array}{c}
\Delta N_5 = w_1 \, \Delta N_1 + w_2 \, \Delta N_2 \: , \\
\Delta N_6 = w_3 \, \Delta N_3 + w_4 \, \Delta N_4 \: .
\end{array}
\end{equation}
\item Calculate the desired value of local column density at $e$ through linear interpolation of $\Delta N_5$ and $\Delta N_6$:
\begin{equation}
\Delta N_e = w_5 \, \Delta N_5 + w_6 \, \Delta N_6 \: .
\end{equation}
\end{list}

Our choice of using the values of local column density at $c_5$ and $c_6$ to arrive at $\Delta N_e$ is arbitrary.
Instead, one may also use the ones from $c_7$ and $c_8$ (cf. Fig. \ref{fig:accumulate3D}) in the steps described above.

The main difficulty in finding an interpolation scheme for the three-dimensional case lies in the fact that we need to weigh with the {\em lengths} of the ray sections to avoid the errors which will otherwise occur when the ray under consideration enters the patch close to a patch corner.
Since in general all these path lengths are different from one another, this introduces quite a number of independent variables into the equations.
So, although the two-step procedure just described is not unique, it is simple and fast, and it gives good results in practice.
%
%
\subsection{Summary of the algorithm}
\label{sect:algoSum}
The steps taken in the algorithm can be summarized as follows (see Fig. \ref{fig:summary}):
\newcounter{Lcount3}
\begin{list}{\arabic{Lcount3}}
{\usecounter{Lcount3}}
\item Each processor checks if its sub-domain contains the source.
The processor that owns the source stores its patch and processor id and makes it available to all other processors (broadcast).
Note that this id may change during a simulation due to changes in refinement and the consequent redistribition of patches among processors.
\item On each processor, create the local patch-mapping and communicate (reduce) it so that each processor ends up with the global patch-mapping (Sect. \ref{sect:cutList}).
\item On each processor, calculate local column density contributions $\Delta N$ for each patch using the `fast voxel traversal algorithm' (Sect. \ref{sect:rayTraceLocal} and Fig. \ref{fig:summary} left).
\item Communicate (gather) all $\Delta N$ values at patch faces located farthest away from the source (Fig. \ref{fig:summary} centre).
Also, the coordinates and refinement levels of all patches need to be gathered.
This communication is done most efficiently when this information is combined in a single data type of size $S$ (Sect. \ref{sect:communicate}).
\item On each processor, construct for each ray the list of patches that are traversed by that ray (Sect. \ref{sect:cutList}).
\item On each processor, interpolate and accumulate the local contributions $\Delta N$ from the patches that are in the list to arrive at the total column density $N$ (cf. Sect. \ref{sect:accumulate} and Fig. \ref{fig:summary} right) .
\end{list}
%
%
\subsection{Comparison to other methods}
\label{sect:methodCompare}
To conclude this section, we compare our method to two more recent ones that either use some form of adaptivity to trace rays \citep{2005ApJ...618..744J}, or that are parallelized for distributed memory architectures \citep{2005AstrophHeinemann}.
Unlike ours, these methods are intended to solve for the full radiation field, and therefore need to employ multiple sets of rays to sample the angular parameter space.
Depending on the adopted form of the source function, (lambda-)iteration is to be performed as well in order to obtain a converging solution.

\citet{2005ApJ...618..744J} proposed a ray tracing method for cell based AMR intended to be used in calculations of line emission.
Their method uses sets of parallel (in the geometrical sense) long characteristics to find the intensity at cell faces, which are then interpolated to get the intensity at the cell centre using a short characteristic.
This is repeated for a number of directions after which angle averaged quantities are obtained.
This process is then lambda-iterated to get converging line intensities.

Since their method refines on a cell-by-cell basis, and ours employs patches structured in an oct-tree hierarchy, there is a one-to-one correspondence between the procedures of ray tracing used in the two methods: their long characteristics correspond to our ray tracing of the patch-mapping, whereas their short characteristics correspond to our ray tracing of a single patch.

More recently, \citet{2005AstrophHeinemann} developed a method for tracing rays through a decomposed computational domain (i.e. sub-domains that are distributed over a number of processors).
To sample the radiation field, rays are traced that are either parallel or diagonal to a regular patch.
As they mention, this means that there is no need for them to interpolate local values.
Furthermore, since their source function acts only locally, their is no need to iterate the solution.

As in our approach, \citet{2005AstrophHeinemann} first obtain all local contributions (which they call `intrinsic') and add these up to arrive at the total solution.
However, in contrast to our method, the communication pattern of \citet{2005AstrophHeinemann} is intrinsically serial (i.e. processors have to wait for one another, see their Fig. 1).
In their case of a decomposed regular domain, the performance penalty due to the serial nature of their algorithm is small, but in case of an AMR type of grid, the performance would be severely degraded.
\citet{2005AstrophHeinemann} also consider the special case of periodic boundary conditions with rays running only parallel along a coordinate axis.
In such a situation, the boundary values are broadcasted and the inter-processor communication is more efficient than in the serial case.

Although our method is designed to study the effects of ionization due to point sources of radiation, it can be easily adapted to trace sets of parallel rays instead.
Depending on the application, a prescription for the source function and \mbox{(lambda-)iteration} would need to be implemented.
This would make our method suitable for solving the radiative transfer equation in a more general way, similar to the methods just discussed.
The added advantage of such an approach is that our method is highly parallel {\em and} coupled to an AMR hydrodynamics code.
%
%
\section{Ionization, heating, cooling}
\label{sect:ionizeHeatCool}
When the column density from the source up to each cell face is known, the ionization fractions and temperature can be computed.
For this we use a simplified version of the DORIC routines \citep[see][]{2002A&A...394..901M,1994A&A...289..937F}.
In what follows, we summarize the way in which these routines calculate the ionization, heating, and cooling rates (for more details, please refer to \citet{1994A&A...289..937F}).
Although the DORIC package is capable of handling a large number of species (H, He, C, N, O, and Ne), we use hydrogen as the only gas component in order to keep the complexity of our tests cases at a minimum, and we will therefore describe just this case.

The ionization fractions of hydrogen are given by
\begin{equation}
  x(\mathrm{HI}) = \frac{n(\mathrm{HI})}{n(\mathrm{H})} , \;
  x(\mathrm{HII}) = \frac{n(\mathrm{HII})}{n(\mathrm{H})} ,
\end{equation}
with
\begin{equation}
  n(\mathrm{H})=n(\mathrm{HI})+n(\mathrm{HII})
\end{equation}
the total hydrogen number density.
The electron number density follows from
\begin{equation}
  n_\mathrm{e} = n(\mathrm{HII}) + n(\mathrm{C}) ,
\end{equation}
where the number density of carbon is included to prevent the possibility of $n_\mathrm{e}=0$, by assuming that carbon is always at least singly ionized due to the interstellar UV field.

For hydrogen, the number of photoionizations per second is given by \citep{1989agna.book.....O}
\begin{equation}
\label{eq:photrate}
  A_\mathrm{p}=\int_{\nu_0}^{\infty}\frac{4\pi J_\nu}{h\nu} a_0 \, d\nu ,
\end{equation}
with $J_\nu$ the local mean intensity of the radiation field, $a_0=6.3\times 10^{-18}\:\mathrm{cm}^2$ the cross section (which we take to be frequency independent, or `grey', for simplicity) and $\nu_0$ the ionization threshold frequency.

The number of collisional ionizations per second is calculated using
\begin{equation}
  A_\mathrm{c} = A_\mathrm{c}(\mathrm{HI}) n_\mathrm{e} \sqrt{T} \exp(-I(\mathrm{HI})/kT)
\end{equation}
with $A_\mathrm{c}(\mathrm{HI})=5.84\times 10^{-11}\:\mathrm{cm}^3\mathrm{K}^{-1/2}$, and $I(\mathrm{HI})$ the hydrogen ionization potential  \citep{1970PhDT.........2C}.

For the on-the-spot approximation, the radiative recombination rate is given by \citep[cf.][]{1989agna.book.....O}
\begin{equation}
  \alpha_\mathrm{R} = \alpha_\mathrm{R}(10^4\mathrm{K})\left(\frac{T}{10^4}\right)^{-0.7} ,
\end{equation}
with $\alpha_\mathrm{R}(10^4\mathrm{K})=2.59\times 10^{-13}\:\mathrm{cm}^3\mathrm{s}^{-1}$.
The temperature is determined from the pressure using
\begin{equation}
  p=(n(\mathrm{H})+n_\mathrm{e})kT .
\end{equation}

The rate equation for the hydrogen ionization fraction is given by
\begin{equation}
  \frac{\mathrm{d} x(\mathrm{HII})}{\mathrm{d}t} = x(\mathrm{HI}) A - x(\mathrm{HII})n_\mathrm{e}\alpha_\mathrm{R} ,
\end{equation}
with $A=A_\mathrm{p}+A_\mathrm{c}$ the total number of photo- and collisional ionizations per second.

When one assumes that the electron density is constant, an analytic solution for $x(\mathrm{HII})$ can be found, and iterating for $n_\mathrm{e}$ gives the time dependent solution \citep{1987A&A...174..211S}.
Since $A_c$ and $\alpha_\mathrm{R}$ are both temperature dependent, the change in temperature due to heating and cooling needs to be recalculated for each iteration step as well.

The photoionization heating rate is given by
\begin{equation}
\label{eq:photheatrate}
  \Gamma_\mathrm{p} = n(\mathrm{HII}) \int_{\nu_0}^{\infty} \frac{4\pi J_\nu}{h\nu} a_0 h(\nu-\nu_0)\, d\nu ,
\end{equation}
and for the cooling rate we use a collisional equilibrium cooling curve from \citet{1972ARA&A..10..375D} (more general composition-dependent cooling is available in the DORIC package).

For calculating the local mean intensity of the radiation field we use a blackbody spectrum, so we have
\begin{equation}
  4\pi J_\nu(\boldsymbol{r}) = \left(\frac{R_\mathrm{S}}{|\boldsymbol{r}|}\right)^2 \frac{2\pi}{c^2} \frac{h\nu^3}{\exp(\frac{h\nu}{k T_\mathrm{S}})-1} \exp(-\tau(\boldsymbol{r})) .
\end{equation}
Here, $R_\mathrm{S}$ is the radius, and $T_\mathrm{S}$ is the effective temperature of the source.
The optical depth $\tau$ at position $\boldsymbol{r}$ of equation (\ref{eq:opticaldepth}) is calculated using the total column density $N(\boldsymbol{r})$ from equation (\ref{eq:totcoldens}).
Since evaluating the integrals for the photoionization and heating rate [equations (\ref{eq:photrate}) and (\ref{eq:photheatrate})] is too time consuming to perform for every value of $\tau$, they are stored in look-up tables for a range of optical depths and interpolated when needed.

The hydrodynamics and ionization calculations are coupled through operator splitting.
To avoid having to take time steps that are the minimum of the hydrodynamics, ionization, and heating/cooling time scales, we use the fact that the equations for the ionization and heating/cooling can be iterated to convergence.
Since these are so called `stiff' equations \citep[e.g.][]{1992nrfa.book.....P}, we use a special iteration scheme \citep{1994A&A...289..937F}.
This means that the only restriction on the time step comes from the hydrodynamics (i.e. the Courant condition).
See \citet{1994A&A...289..937F} for an assessment of the validity of this approach.
%
%
\section{Tests}
\label{sect:testing}
In this section we present a number of tests for our new algorithm.
First, we discuss the accuracy with which column densities and ionization fractions are calculated.
We compare the results obtained with the hybrid characteristics method to those calculated using long characteristics.
Since the interpolation scheme of our method is designed to give the exact result for the column density in case of a homogeneous density distribution (Sect. \ref{sect:accumulate}), we also consider its performance in case of a more general density field.

We conclude this section by testing the shadow casting capabilities of our method and apply it to a `real-world' application of photoevaporating flows.
This last test demonstrates the performance of out method when used in combination with hydrodynamics.
%
%
\subsection{Column density}
\label{sect:coldenstest}
%
%
\begin{figure*}
\centering
  \includegraphics[width=17cm]{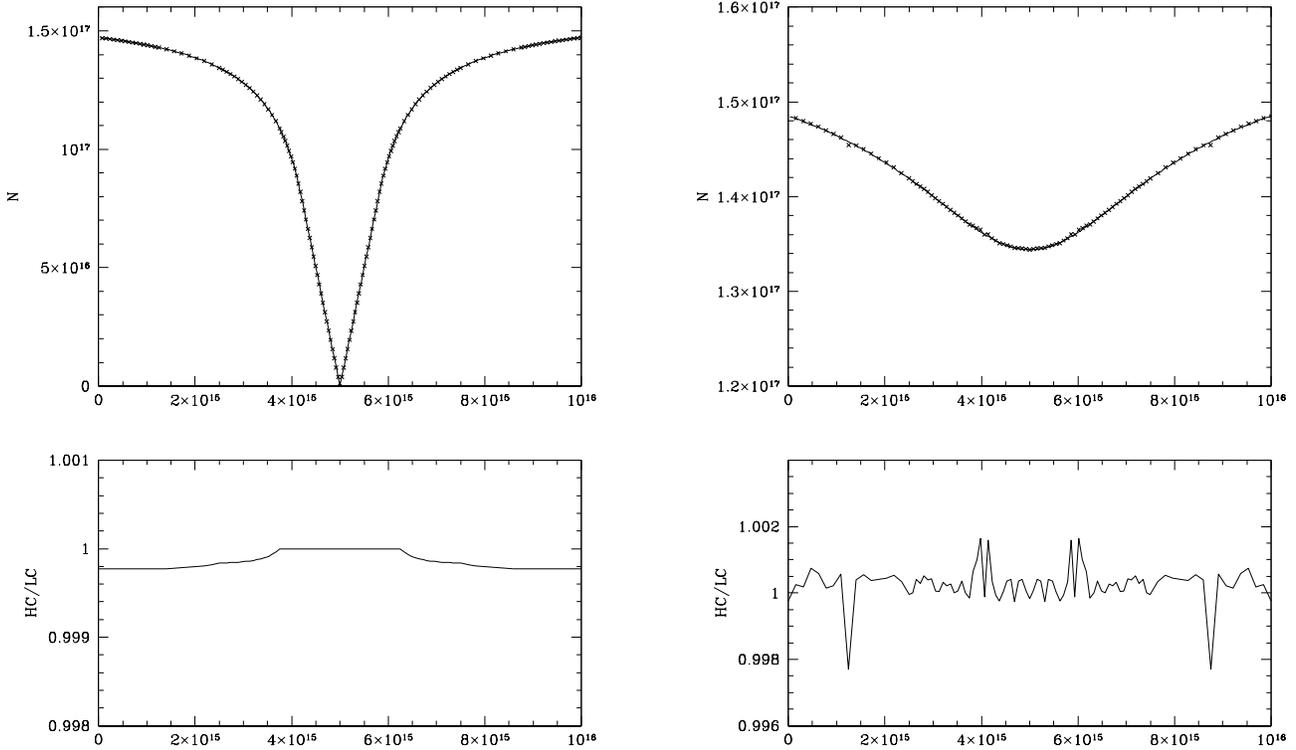}
  \caption{
Values of column density for the case of a single point source in a two-dimensional domain with a $1/r^2$ density distribution.
Shown are one-dimensional cuts along the y-direction through the source located at the centre of the domain (left panels) and at $3/4$ of the domain (right panels).
In the top panels, the solid line indicates the result for the long, whereas the crosses indicate the result for the hybrid characteristics method.
The bottom panels show the ratio (hybrid/long) of column density values.
}
  \label{fig:columnTestR2}
\end{figure*}
%
%
We performed two-dimensional calculations where we placed a single point source at the centre of a $1/r^2$ density distribution, the result of which is shown in Fig. \ref{fig:columnTestR2}.
In order to prevent an under-resolved singularity at the location of the source, we used a constant density sphere with a radius of $5\times 10^{14}$ at the source location.

The left panel of Fig. \ref{fig:columnTestR2} shows the column density distribution along a line $y=\mathrm{const}$ cutting through the exact location of the source.
Since for this special case no interpolation is necessary, only very small differences between the two methods are found.
These differences are due to uneven sampling of the $1/r^2$ density distribution on the adaptive mesh.
The errors increase only slightly ($< 0.5\%$) when interpolation is used, as indicated by the results obtained along the $y=\mathrm{const}$ line located at $3/4$ of the horizontal extent of the domain (the right panel in Fig. \ref{fig:columnTestR2}).
%
%
\subsection{Shadow casting}
\label{sect:shadowtest}
To test the shadow casting capabilities of our algorithm, we calculate the column density and ionization fractions for a homogeneous environment containing higher density clumps, which are taken to be spherical and neutral.
The ionization state is found by iterating the ionization fractions over a period equal to a few recombination time scales, while keeping the temperature fixed.

The computational domain spans the region $(2.0,1.0,1.0)\times 10^{18}\:\mathrm{cm}$.
The ambient medium has a number density $n_\mathrm{env}=10^2\:\mathrm{cm}^{-3}$ and a temperature $T_\mathrm{env}=5000\:\mathrm{K}$.
The source of ionizing radiation is located at $(x,y,z)=(0.0,0.5,0.5)\times 10^{18}\:\mathrm{cm}$.
It has a luminosity $L_\mathrm{S}=7000\:L_\odot$ and an effective temperature $T_\mathrm{eff}=50000\:\mathrm{K}$.
The resulting Str\"omgren sphere has a radius that is larger ($\sim 3\times 10^{18}\:\mathrm{cm}$) than the physical size of the computational domain.
Two identical clumps are placed at a distance of $\sim 10^{18}\:\mathrm{cm}$ from the source.
Each clump has a density $n_\mathrm{clump}=10^4\:\mathrm{cm}^{-3}$, a temperature $T_\mathrm{clump}=100\:\mathrm{K}$, and a radius $r_\mathrm{clump}=4\times 10^{16}\:\mathrm{cm}$.
We used $6$ levels of refinement with patches of $16^3$ cells.
The effective resolution in this test was $1024\times 512^2$ cells.

\begin{figure*}
\centering
  \includegraphics[height=3.4cm]{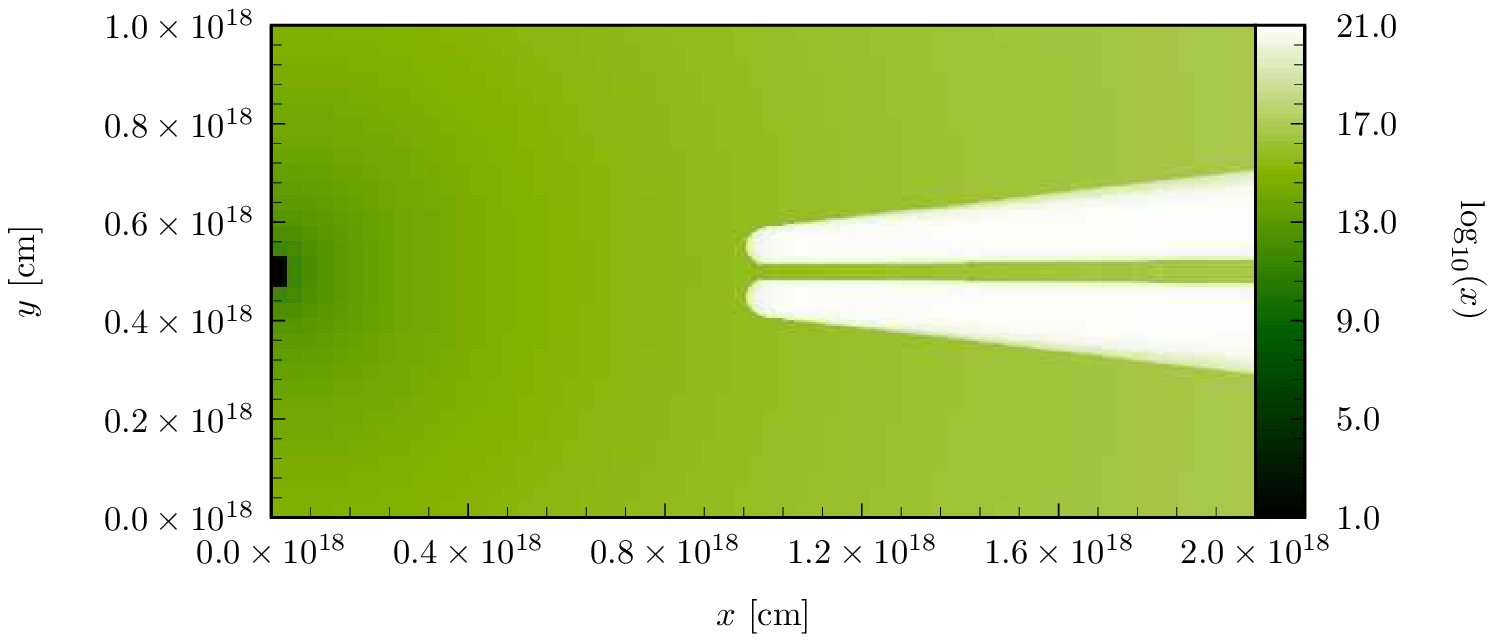}\hspace{0.5cm}
  \includegraphics[height=3.4cm]{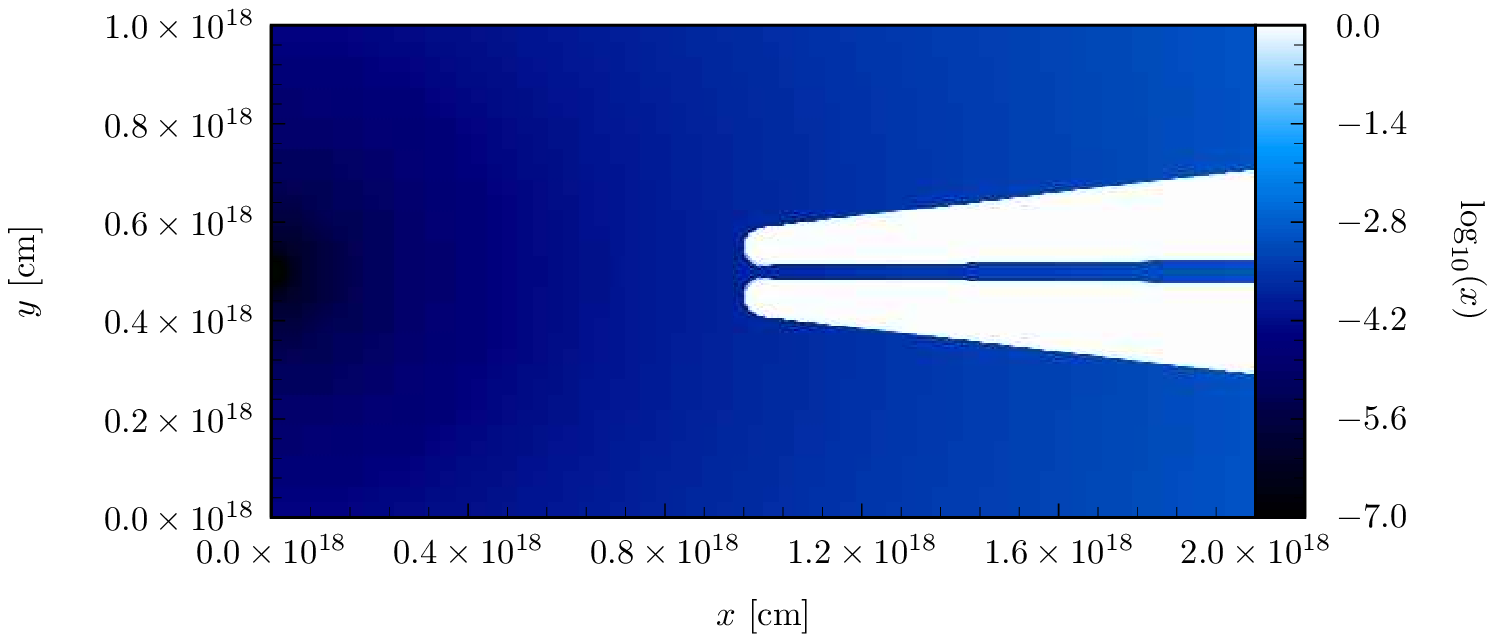}\\
  \vspace{0.5cm}
  \includegraphics[height=3.4cm]{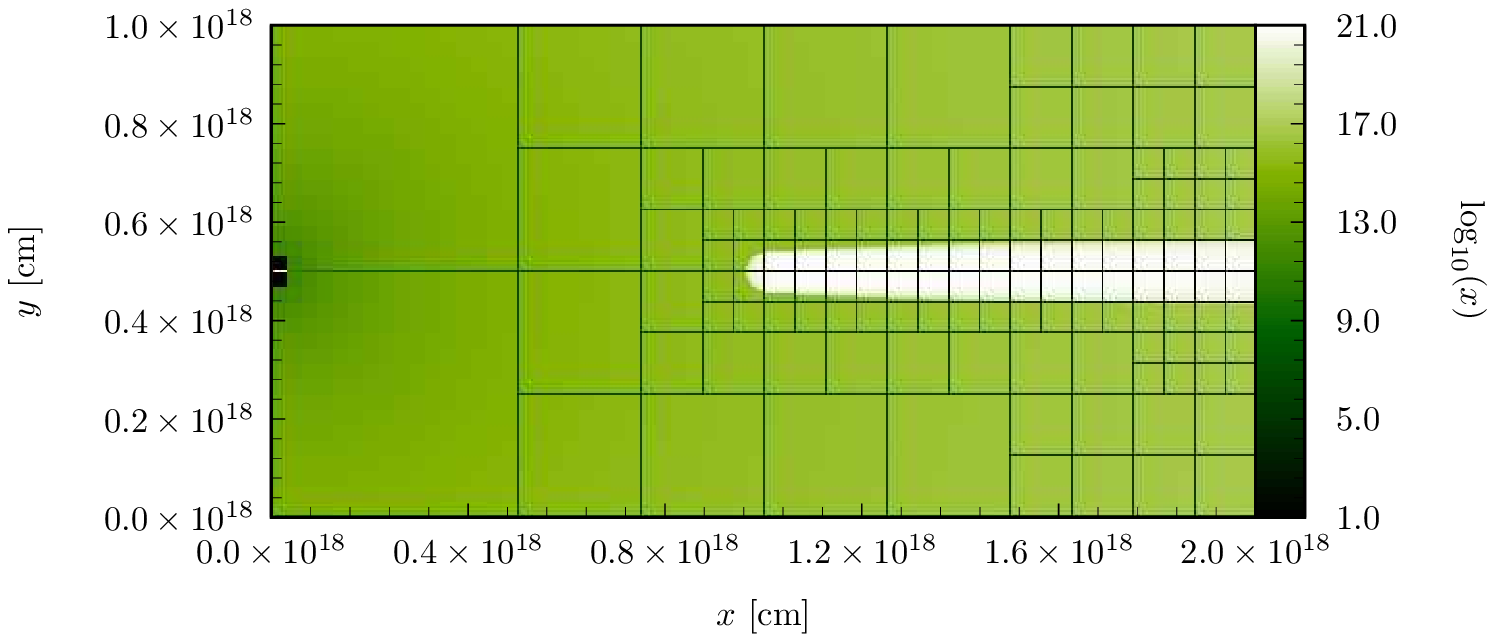}\hspace{0.5cm}
  \includegraphics[height=3.4cm]{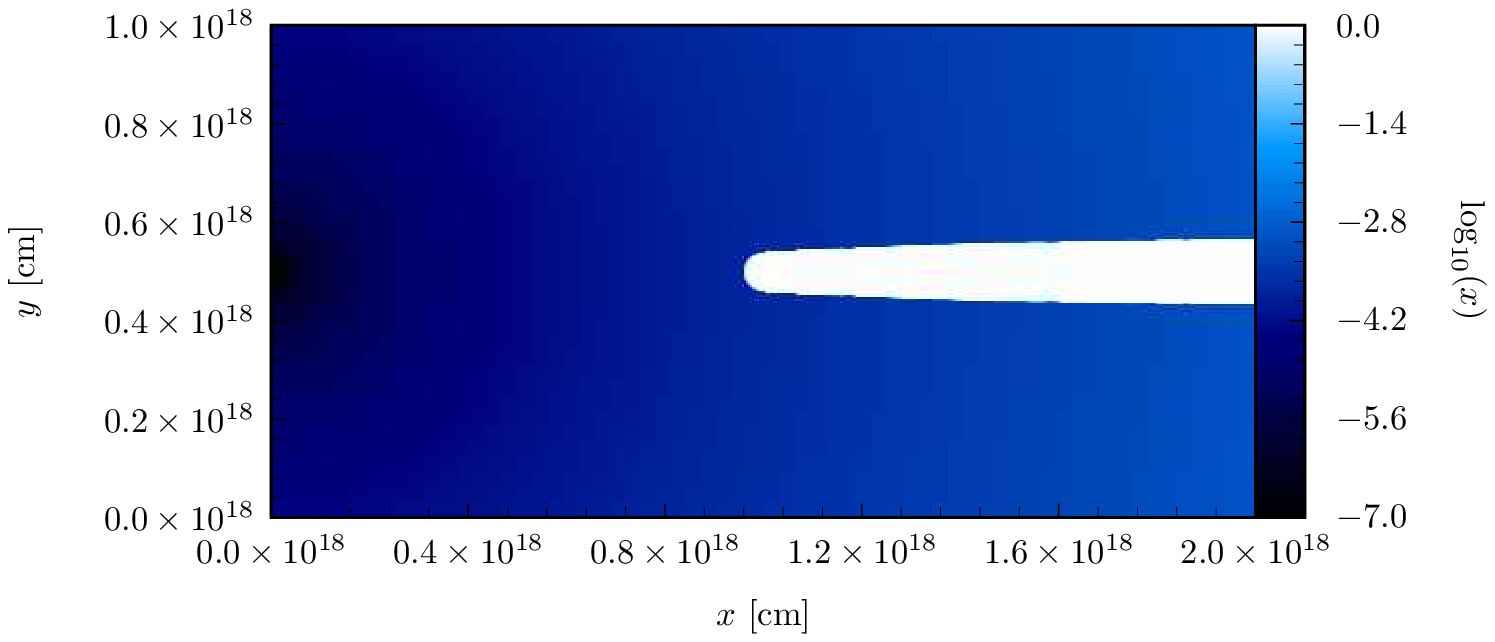}\\
  \caption{
Values of $\log_{10}$ of the column density (left) and $\log_{10}$ of the HI ionization fraction (right) for the case of a single point source in an environment with a homogeneous density distribution containing neutral clumps with higher density.
Shown are color coded plots of xy-cuts through the centre of the domain (top row) and xz-cuts through the centre of the bottom clump (bottom row).
The bottom left image shows the AMR patch distribution superimposed, where each patch contains $16^3$ cells.
}
  \label{fig:shadowTest}
\end{figure*}

The results of the shadow casting test are shown in Fig. \ref{fig:shadowTest}.
As one can see, our hybrid characteristics method is capable of casting shadows with very sharp boundaries, indicating a low numerical diffusivity of the scheme.
We note that since the initial conditions do not contain any density gradient, column densities calculated in this test are identical to the ones one would obtain using a long characteristics method.
%
%
\subsection{Application: photoevaporating clumps}
\label{sect:applications}
To illustrate that our hybrid radiative transfer algorithm can be used efficiently in combination with hydrodynamics, we present a first 3D application of the evolution of over-dense clumps being photoevaporated.
We use the parameters of the simulation setup described in Sect. \ref{sect:shadowtest} as initial conditions and follow the dynamical evolution for $\sim 4000\:\mathrm{yr}$.
This simulation is similar to the ones presented by \citet{2003A&A...405..189L}, with this difference that in our simulation both the source and the clumps are inside the computational domain, and that our radiation field is not approximated by parallel rays.

These computations are relevant to the shaping and evolution of cometary knots which are observed in objects like the Helix (NGC~7293), Eskimo (NGC~2392), and Dumbbell (M27) nebula.
Another application is the interaction zone that is observed between binary proplyds in HII regions like NGC~3603 \citep{2000AJ....119..292B} and the Orion Nebula \citep{2002ApJ...570..222G}.

\begin{figure*}
\centering
  \includegraphics[height=3.4cm]{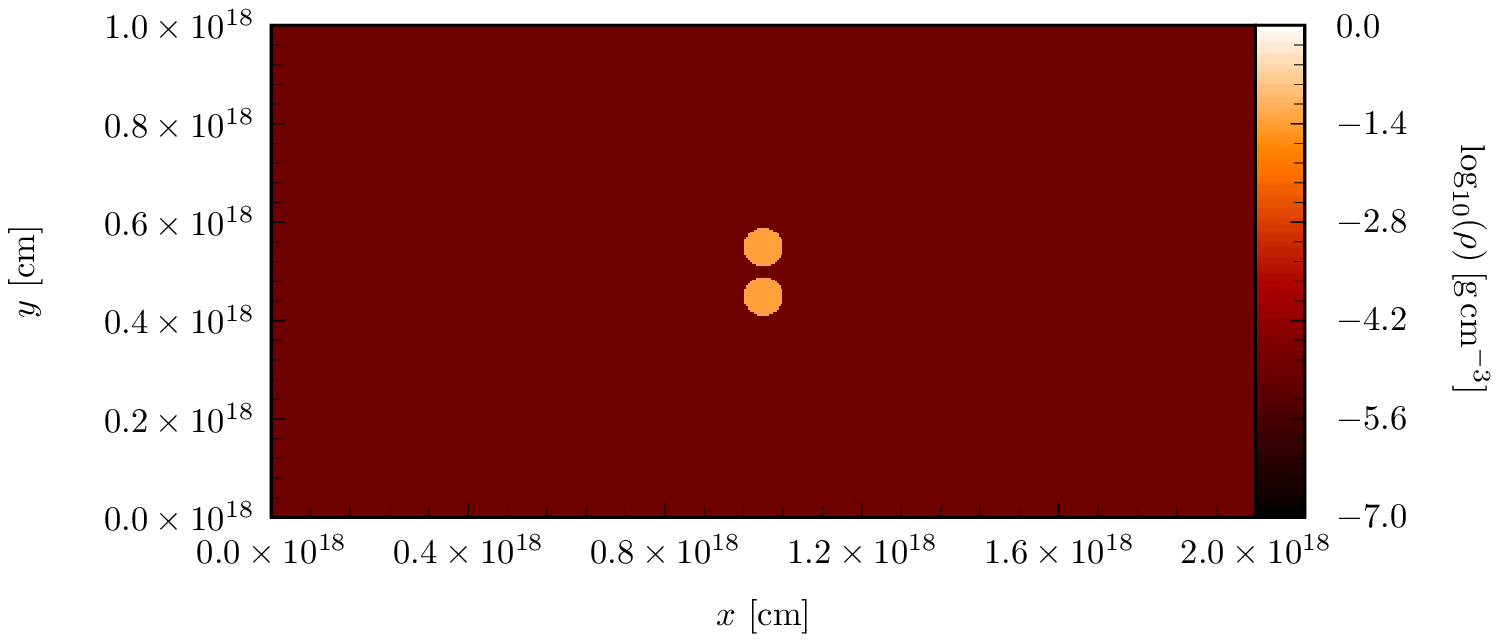}\hspace{0.5cm}
  \includegraphics[height=3.4cm]{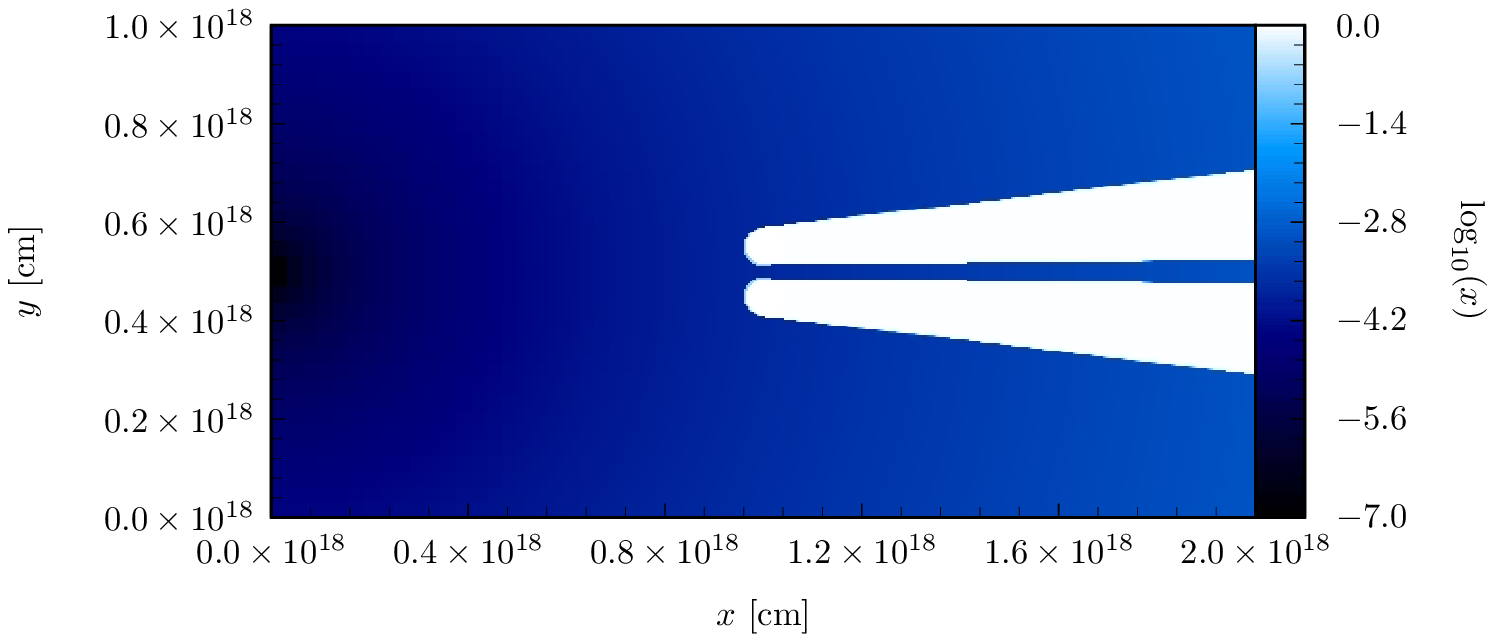}\\
  \vspace{0.5cm}
  \includegraphics[height=3.4cm]{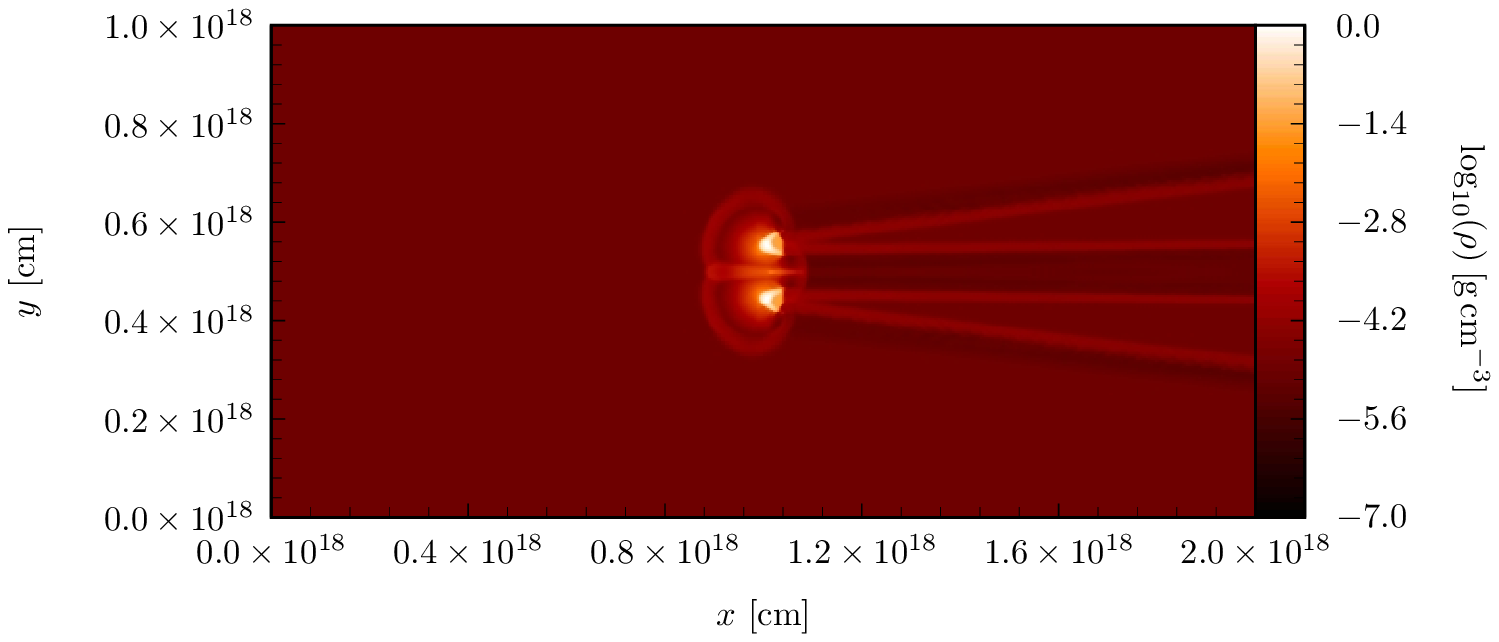}\hspace{0.5cm}
  \includegraphics[height=3.4cm]{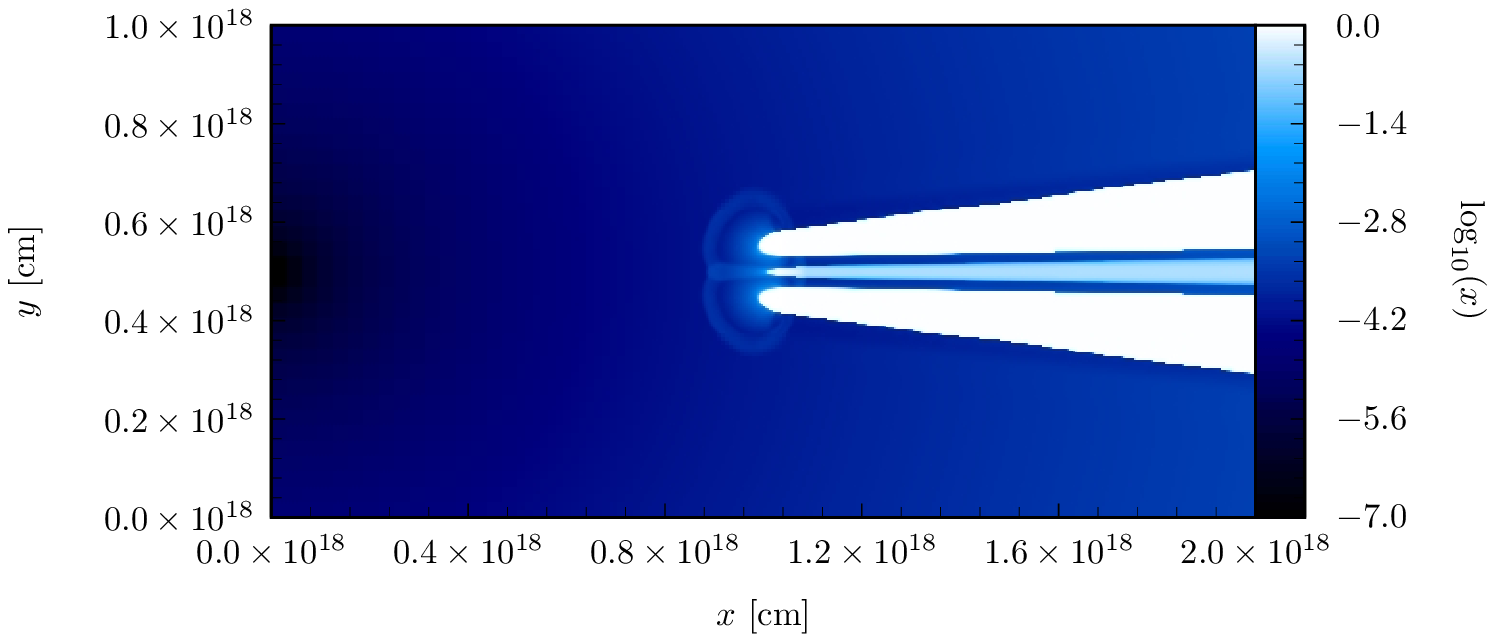}\\
  \vspace{0.5cm}
  \includegraphics[height=3.4cm]{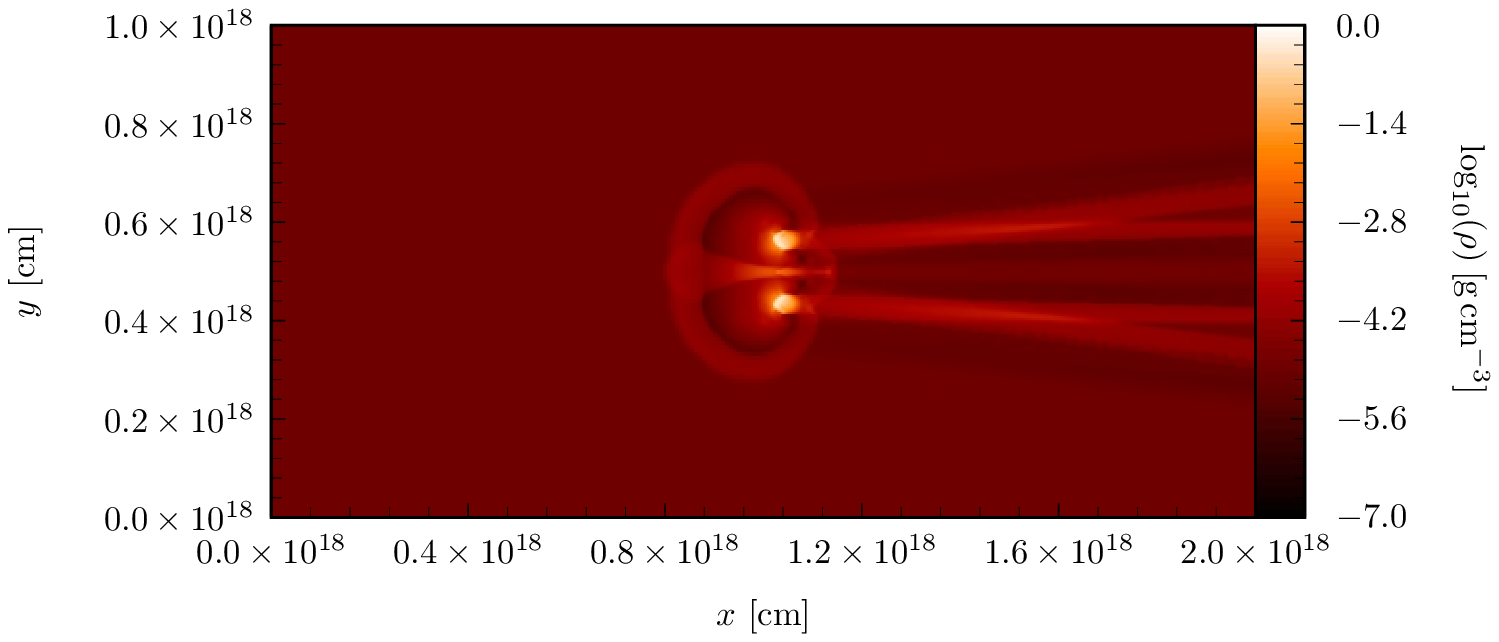}\hspace{0.5cm}
  \includegraphics[height=3.4cm]{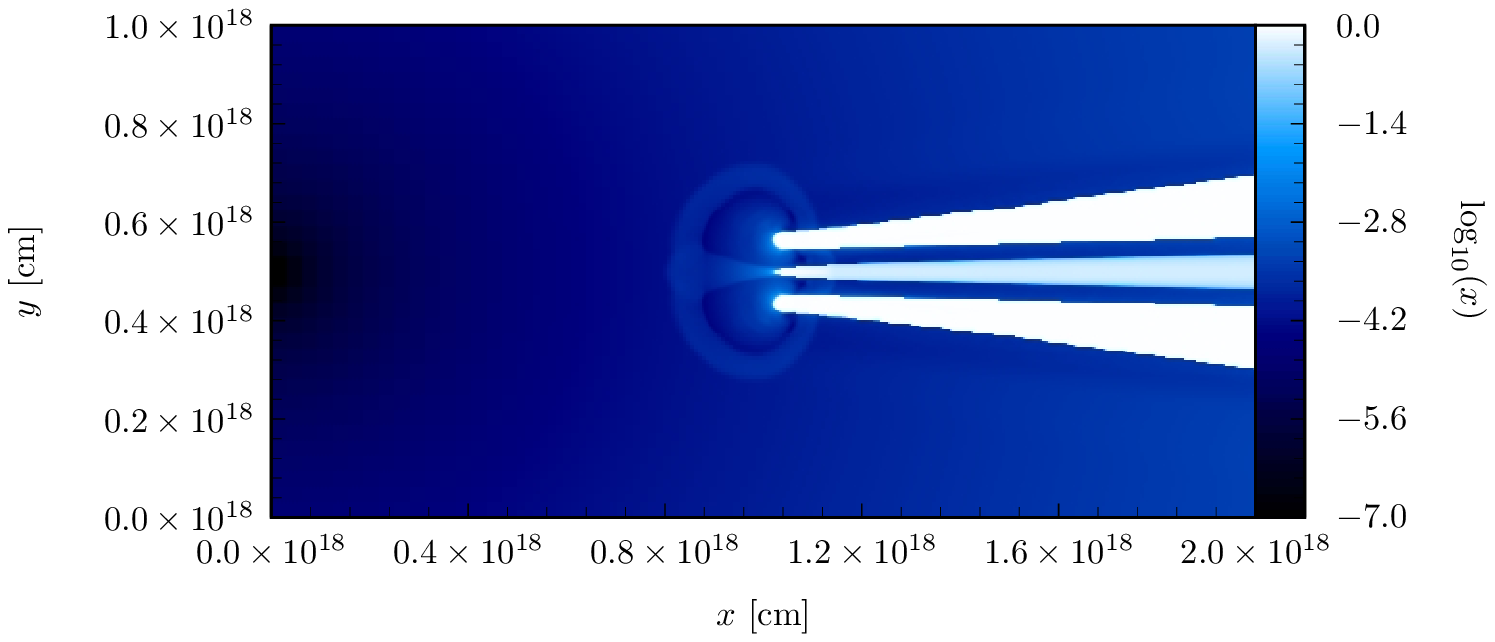}\\
  \vspace{0.5cm}
  \includegraphics[height=3.4cm]{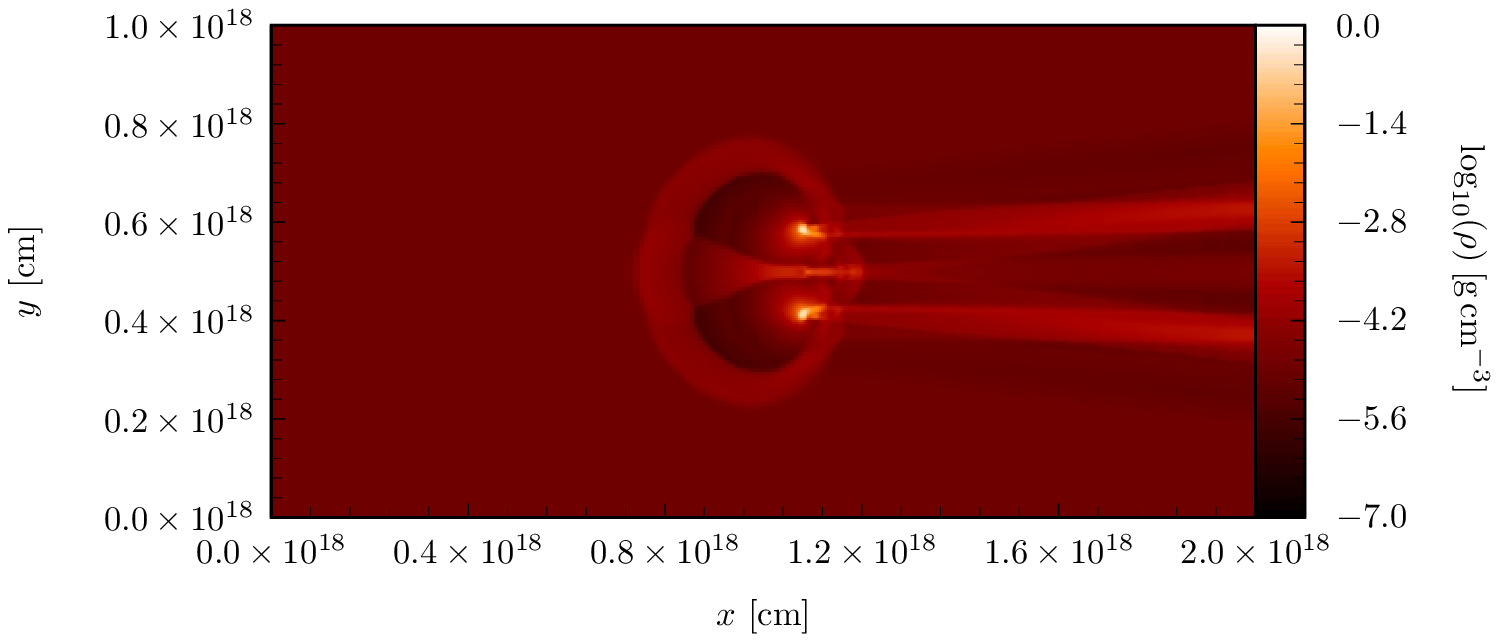}\hspace{0.5cm}
  \includegraphics[height=3.4cm]{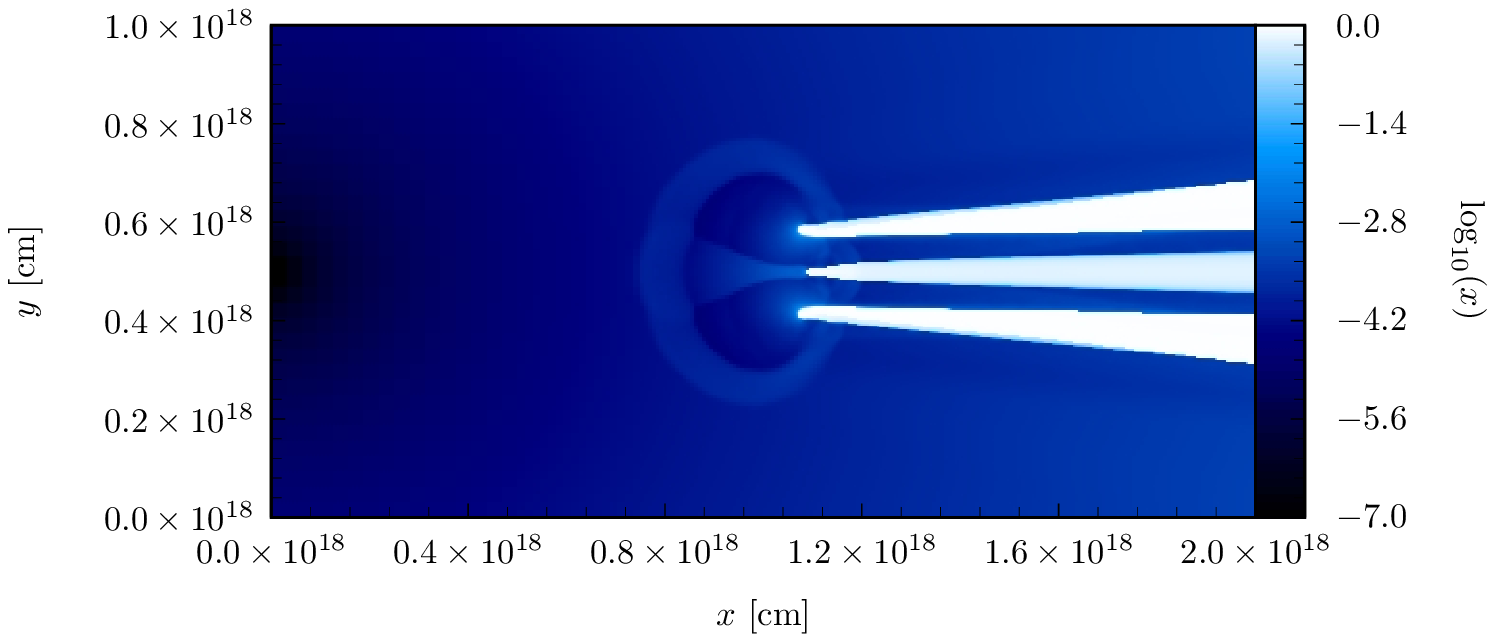}\\
  \vspace{0.5cm}
  \includegraphics[height=3.4cm]{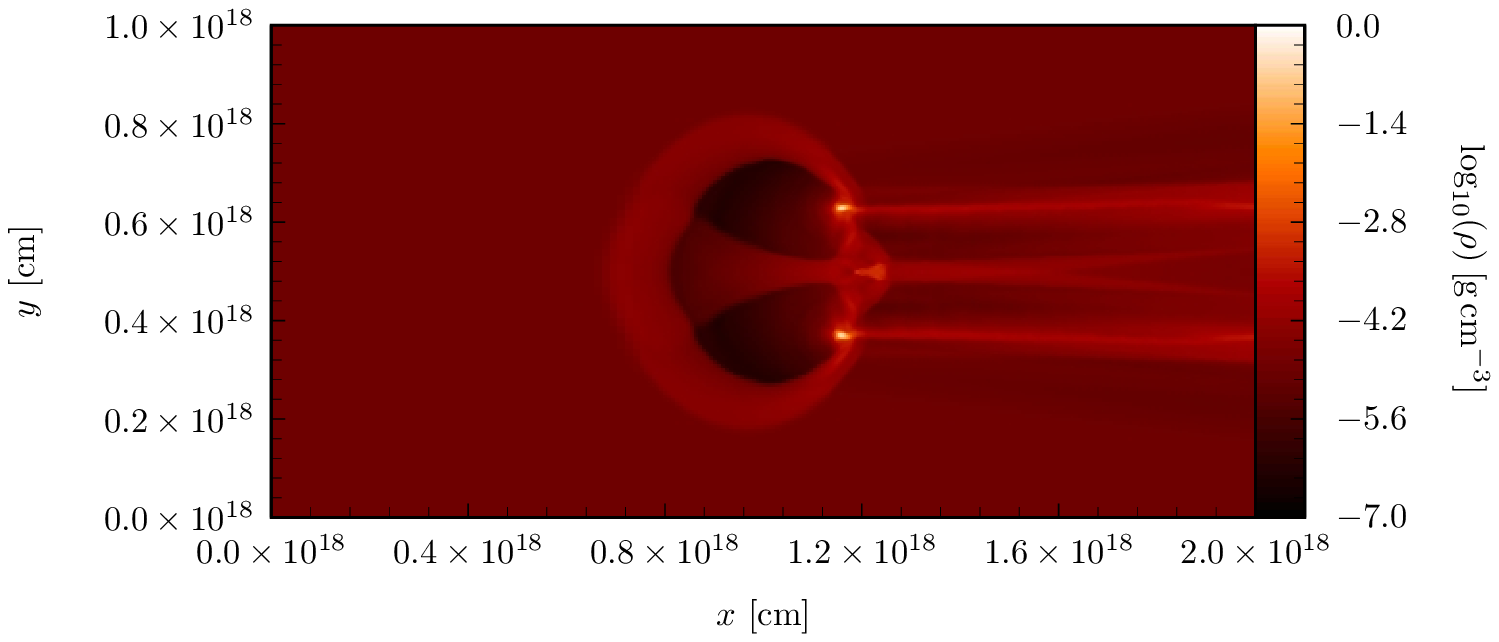}\hspace{0.5cm}
  \includegraphics[height=3.4cm]{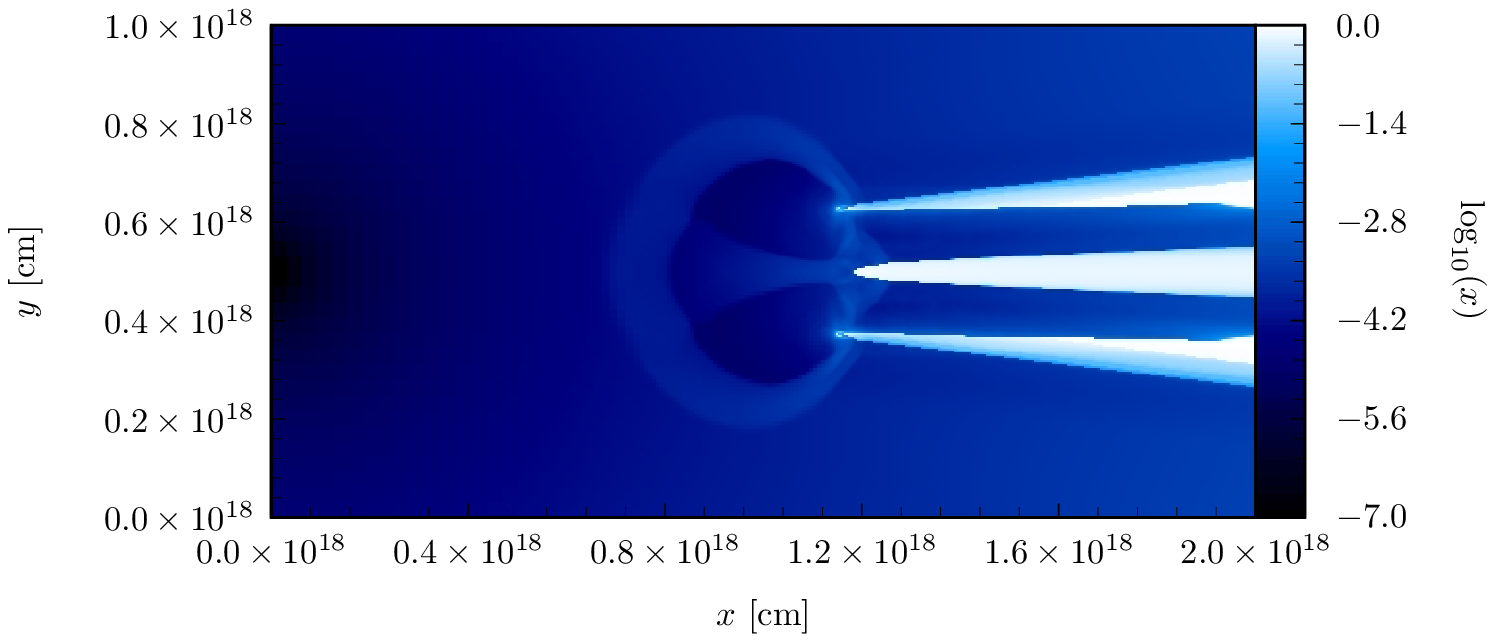}\\
  \vspace{0.5cm}
  \includegraphics[height=3.4cm]{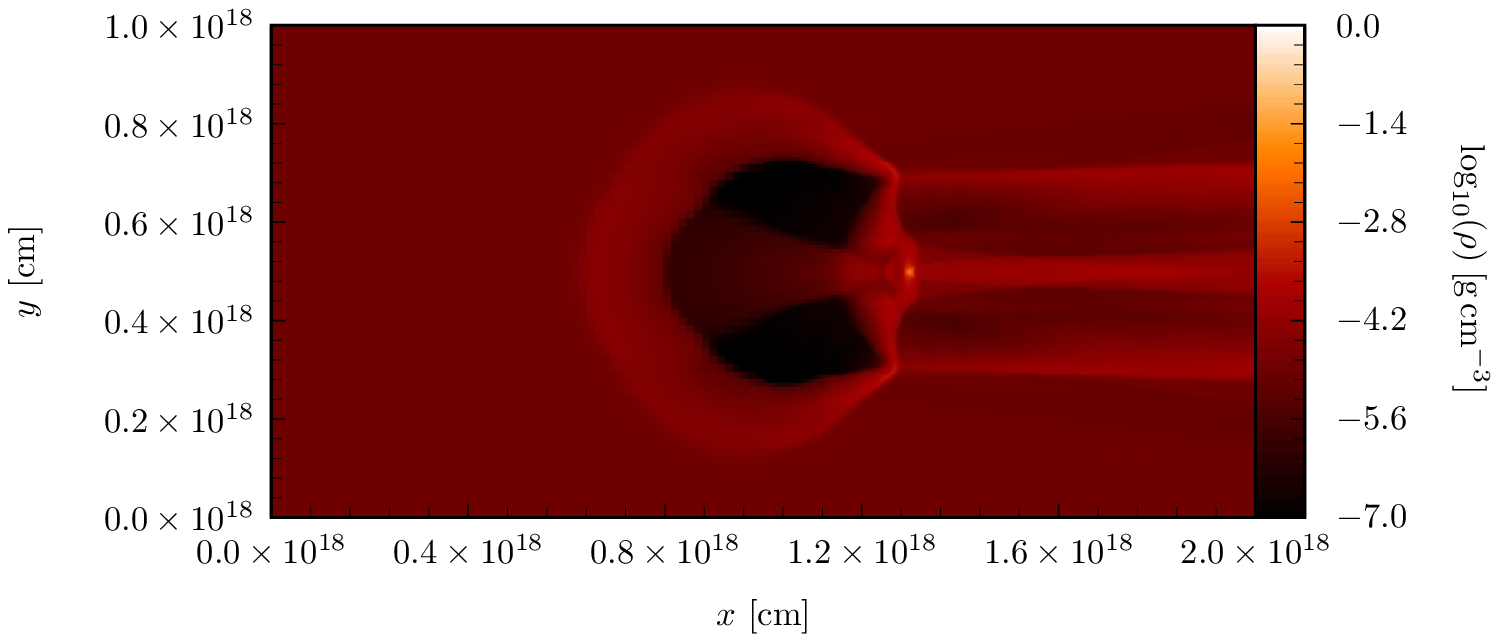}\hspace{0.5cm}
  \includegraphics[height=3.4cm]{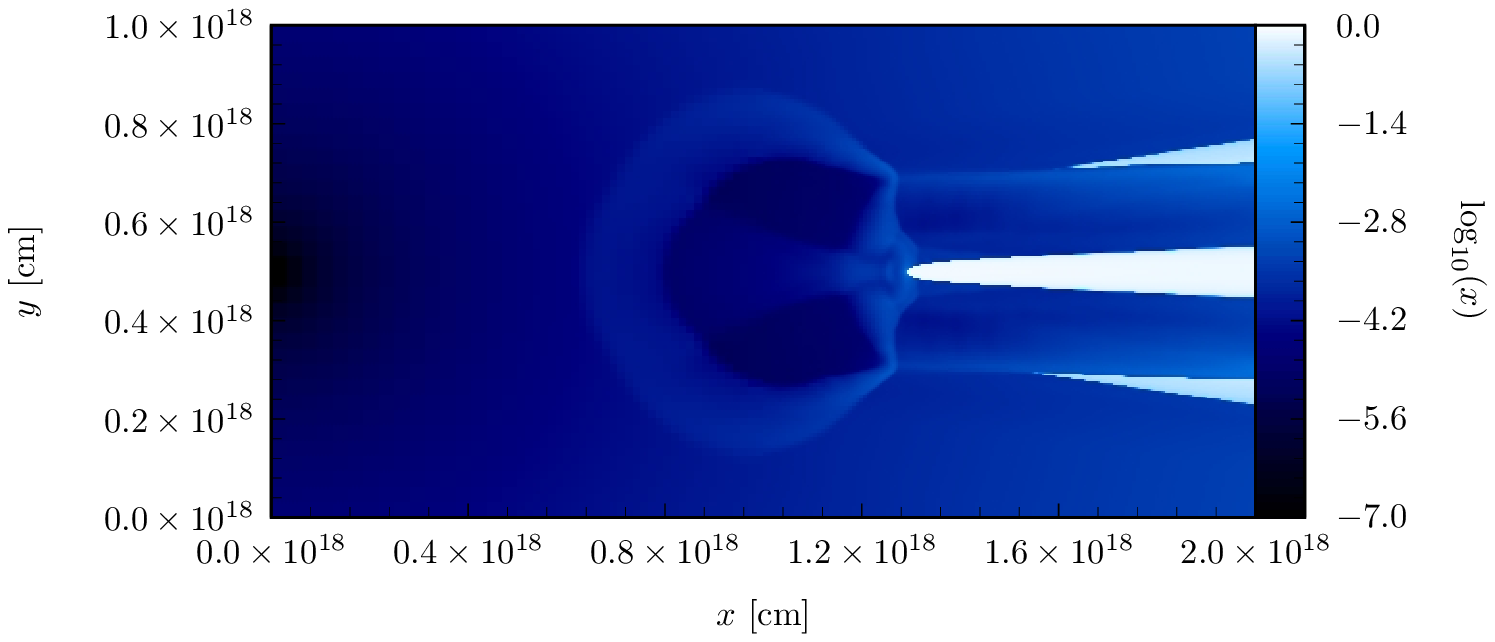}\\
  \caption{
Snapshots of the evolution of the $\log_{10}$ of the mass density (left) and the $\log_{10}$ of the HI ionization fraction (right) for the case of a single point source in an environment with a homogeneous density distribution containing two neutral clumps with higher density.
The source is located at $(0.0,0.5,0.5)\times 10^{18}\:\mathrm{cm}$.
Shown are color coded plots of xy-cuts through the centre of the domain at $t=0\:\mathrm{yr}$ (first row), $t=792\:\mathrm{yr}$ (second row), $t=1584\:\mathrm{yr}$ (third row), $t=2377\:\mathrm{yr}$ (fourth row), $t=3169\:\mathrm{yr}$ (fifth row), and $t=3961\:\mathrm{yr}$ (sixth row).
}
  \label{fig:clumpCuts}
\end{figure*}

Fig. \ref{fig:clumpCuts} shows a sequence of snapshots \footnote{Movies of the simulations are available with the electronic version of this paper at \texttt{http://www.edpsciences.org}} of the density and neutral hydrogen fraction at different times during the simulation.
One sees that the interaction of the photoevaporation flows coming from the clumps results in a zone of higher density between the clumps, which, as was already found by \citet{2003A&A...405..189L}, can explain the excess emission observed between some cometary knots and binary proplyds.
This interaction zone recombines, becomes optically thick, and casts a shadow.
It is interesting to see that the zone, and the shadow region behind it, persist even after the two clumps have been fully evaporated.
This mechanism for creating extra shadows may influence the evolution and survival time of clumps that lie farther away from the star, an effect not taken into account in previous numerical studies.
We are currently investigating further into this kind of flows and will present our findings in a forthcoming publication.
%
%
%
%
%
%
%
%
%
%
\section{Performance analysis}
\label{sect:performance}
We start this section by comparing our hybrid characteristics method to the more traditional long and short characteristics methods for regular grids.
In order to do this, we distinguish between two types of computations: first we determine how many calculations are needed to arrive at the local contribution each cell makes to the column density along a ray, and second we look at the number of interpolations the different methods have to perform to compute the total column density up to each cell.

Consider a computational domain with a resolution of $C^3$ cells and a source located at one of the corners of the domain.
For the case of a regular grid, the maximum number of cells a long characteristic would encounter is $\sqrt{3}C$, and, since we assume that a ray is cast to all cells, the number of calculations needed to provide the total column density is therefore $\lesssim C^4$.

For our hybrid characteristics method, which employs an oct-tree type of AMR grid, the maximum number of cells a local ray section encounters is $\sqrt{3}c$, where $c^3$ is the number of cells in a single patch (cf. Sect. \ref{sect:rayTraceLocal}).
So in this case, for a fully refined grid, the total number of calculations would amount to $\lesssim c\, C^3$.
But, since in general the domain would be refined by a factor $r$, with $0\leq r \leq 1$, this number reduces to $\lesssim r\, c\, C^3$.
This means that when $r\,c \simeq 1$, our method needs $\sim C^3$ calculations to arrive at the local contributions to the column density for each cell, which is of the same order a short characteristics method would need on a regular grid.
However, the local values of column density still need to be communicated and interpolated to arrive at the total column density for each cell.
On the other hand, a short characteristics method also needs to interpolate local values when it sweeps through the grid, whereas a long characteristics method, although it executes a factor $C$ more calculations, does not need to perform any interpolations at all.

The number of interpolations to be performed by our method is determined by the number of patches that are encountered when ray tracing through the patch-mapping (cf. Sect. \ref{sect:cutList} and \ref{sect:accumulate}).
This number is at most $\sqrt{3}C/c$, since, for a fully refined grid, there are $C/c$ patches along a coordinate axis.
For a grid that is not fully refined this number is again reduced by a factor $r$.
A ray trace through the patch-mapping is to be performed for every cell, which brings the total number of interpolations to $\lesssim r^2 (C/c) C^3$, where one factor of $r$ comes from the number of patches cut by a ray, whereas the other factor comes from the total number of cells that exist in the computational domain.

A short characteristics method needs to do an interpolation for every cell, so, for a regular grid, the total number of interpolations is $C^3$.
This implies that when $r^2 C/c \simeq 1$, our method needs to compute a similar number of interpolations as a short characteristics one.
Note that we assume that the calculations needed to do the interpolations are comparable in execution speed for the short and hybrid characteristics methods, which may actually not be the case.

As an example, a typical AMR calculation has $C=512$, $c=16$ (i.e. $6$ levels of refinement), and $r=0.25$, which results in $rc=4$ and $r^2 C/c = 2$.
This shows that, for a single processor calculation with a proper choice of the ratio $C/c$ and a reasonable amount of refinement, our hybrid characteristics method is expected to perform equally well as a short characteristics method on a regular grid.
It also means that, when our method is used in parallel, a better performance will be obtained when increasing the number of processors.

To investigate this aspect in some more detail we have conducted a preliminary performance analysis using the photoevaporating clumps test case described in Sect. \ref{sect:applications} as the underlying physics problem.
We used $5$ levels of refinement irrespectively of the number of processor used in the test run (i.e., the problem had a fixed total work).
Calculations have been terminated after reaching $10\%$ of the nominal simulation time.
Otherwise the simulation parameters were identical to those used in the calculations presented in Sect. \ref{sect:applications}.

\begin{figure}
\resizebox{\hsize}{!}{\includegraphics[width=10cm,angle=-90]{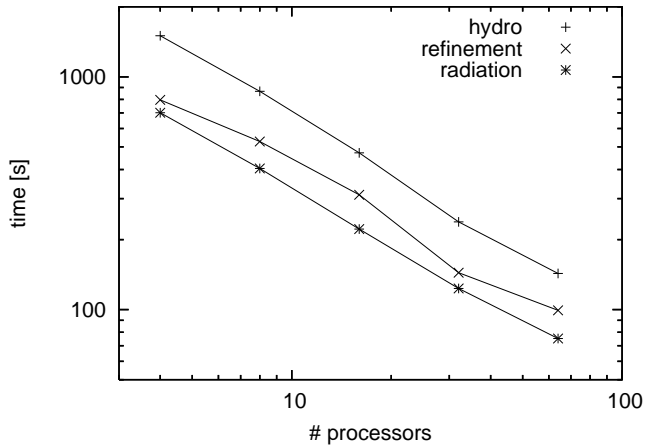}}
  \caption{
Performance of the main components of a radiation hydrodynamics calculation.
}
  \label{fig:overallPerformance}
\end{figure}

\begin{figure}
\resizebox{\hsize}{!}{\includegraphics[width=10cm,angle=-90]{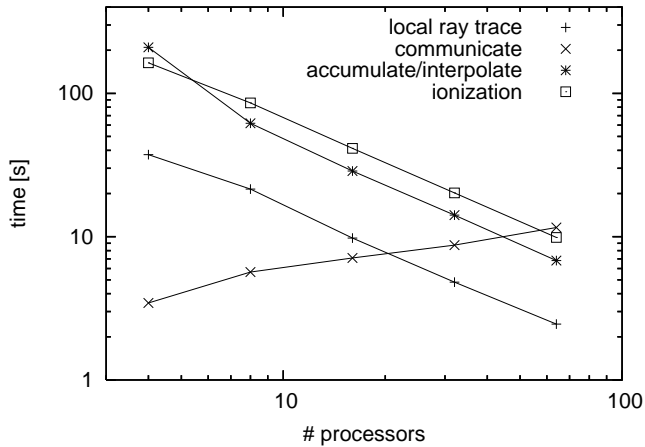}}
  \caption{
Performance of the different steps in our hybrid characteristics method.
For this specific test, the communication takes as much time as the rest of the calculation when using $64$ processors.
As is explained in the text, for patches with a larger number of cells, this constraint may become less severe.
}
  \label{fig:raytracePerformance}
\end{figure}

The results of our performance study are shown in Fig. \ref{fig:overallPerformance}, where we compare the overall performance of the hydrodynamics, AMR, and radiation calculations.
Detailed results for the radiation part are presented in Fig. \ref{fig:raytracePerformance}, where the timings for the individual components of our hybrid characteristics method (local ray trace, communication, accumulation/interpolation, and ionization) are shown.
%
%
%
%
%
%

Performance data obtained for our realistic test problem indicates that the ray tracing part of the calculation takes less time than either the hydrodynamics or grid adaptation.
Furthermore, it shows that most of the computational time during ray tracing is spent in interpolating and accumulating the local contributions to the column density (i.e. step 6, Sect. \ref{sect:algoSum}).
Following the analytical assessment made above, we conclude that in order to reduce the number of interpolations required during calculation one should try to minimize the value of $r^2 C/c$ rather than $rc$ when setting up a simulation.
This suggest that one should use patches that contain a relatively large number of cells compared to the effective resolution of the computational domain, and, of course, keep the filling factor of the finest AMR level at a minimum.

Fig. \ref{fig:raytracePerformance} shows performance results for the radiation module including the ionization package for our fixed size problem.
As one can see, the time required to calculate the column densities is about the same as the time needed to calculate the ionization state of the gas.
Furthermore, both these calculations are local and therefore perform very well.
On the other hand, we notice that the communication part of our algorithm does not perform perfectly.
This is somewhat expected since, with the increasing number of processors, the efficiency of our algorithm becomes limited by the efficiency of the global gather operation (used to collect column densities from patch faces).
The results for this specific test indicate that communication is likely to dominate the runtime when more than $\sim 64$ processors participate in the computations.
We expect that this limitation becomes less severe when larger patches are used in the simulation.
In this case the cost of communication may still be lower than, for example, the time needed to accumulate and interpolate the column densities.
To determine whether this is indeed the case, more elaborate performance tests involving a larger number of processors (of the order of $\sim 1000$) are required, and such tests are currently underway.
%
%
\section{Conclusions}
\label{sect:conclusions}
We described a new radiative transfer algorithm for parallel AMR hydrodynamics codes, called hybrid characteristics.
We presented details of several aspects of the algorithm: ray tracing, communication, and interpolation.

The ray tracing is performed in two steps.
First, local long characteristics are used to calculate column density contributions for each patch.
A second ray trace is then performed where a so called patch-mapping is used to find the patches cut by each ray.
When the list of patches cut by a ray is known, interpolation of local column density values is required to find the total column density up to each cell.
For this, one needs the values of local column density contribution at patch faces, which are communicated to all processors.
The coefficients used in the interpolation are chosen such that the exact solution for the column density is retrieved when there are no gradients in the density distribution.

For the case where the distribution is not constant but has a $1/r^2$ profile we find deviations of the order of $\sim 0.5\%$ when comparing our method with a long characteristics one.
This high accuracy with which column densities values are calculated results in well defined and sharp shadows.

We showed that our method can be used efficiently for parallel radiation hydrodynamics calculations in three dimensions on AMR grids.
We presented preliminary results for our new method in application to the problem of the photoevaporation of two over-dense clumps due to the ionization by a single source of radiation.
The results of this simulation offer a possible explanation for the excess emission observed in between cometary knots seen in for example the Helix Nebula, and the interaction zone observed in binary proplyds found in HII regions like NGC~3603 and the Orion Nebula.
These simulations also suggest a possible mechanism for the creation of extra shadows by the high density interaction zone forming between the clumps.
This additional shadowing may influence the evolution and survival time of clumps that lie farther away from the source.
We are currently investigating further into this kind of interactions between photoevaporating flows, and their consequences for the dynamics, and will report our findings in a future publication.

An initial performance test showed that our method works very well when used for calculations on a parallel machine.
For this specific test, the communication part of our algorithm starts to dominate the calculation when more than $\sim 64$ processors are used.
However, we showed analytically that a careful choice of the ratio of the number of cells per patch to the total number of cells in the computational domain controls the amount of communication used in the calculation.
This analysis can be used to optimize the design of our method.
More in-depth performance and scaling studies are currently underway, using large ($\sim 1000$) number of processors, and these will also be used to further optimize the current implementation.

Because of the modular nature of the FLASH code and the DORIC routines, additional elements like more sophisticated cooling or multiple species can easily be added.
Also, multiple point sources can be handled by our method, and in principle moving sources could be implemented.
Furthermore, it should be straightforward to extend the hybrid characteristics method so that it can be used to solve for a more general radiation field, with the added advantage that our method is already parallelized and coupled to an AMR hydrodynamics code.

Another possible application for our method is the calculation of the propagation of ionization fronts in a cosmological context.
For these calculations photon conservation is an important issue.
Recently, \citet{MellemaInPrep2005} developed a method for following R-type ionization fronts that may move more than one cell per time step, where a special formulation of the equations ensures photon conservation.
Although the parallel nature of our algorithm may complicate the implementation of such an approach, we may still benefit from the ideas presented by \citet{MellemaInPrep2005}.

We intend to make our method publicly available in a future FLASH release.
%
%
\begin{acknowledgements}
EJR wishes to thank the ASC Flash Center for its hospitality during a very enjoyable and productive visit when most of the work presented in this paper was done.

This work is supported in part by the U.S. Department of Energy under Grant No. B523820 to the Center for Astrophysical Thermonuclear Flashes at the University of Chicago.

EJR was sponsored by the National Computing Foundation (NCF) for the use of supercomputer facilities, with financial support from the Netherlands Organization for Scientific Research (NWO), under Grant No. 614.021.016.

The work of GM is partly supported by the Royal Netherlands Academy of Arts and Sciences (KNAW).
\end{acknowledgements}
%
%
\appendix
\section{Fast voxel traversal}
\label{app:voxels}
%
%
\begin{figure}
\psfrag{txi}{$t_x^\mathrm{i}$}
\psfrag{tyi}{$t_y^\mathrm{i}$}
\psfrag{tmax}{$t_\mathrm{max}$}
\psfrag{dtx}{$\delta t_x$}
\psfrag{dty}{$\delta t_y$}
\psfrag{Dx}{$\Delta x$}
\psfrag{Dy}{$\Delta y$}
\resizebox{\hsize}{!}{\includegraphics[width=10cm]{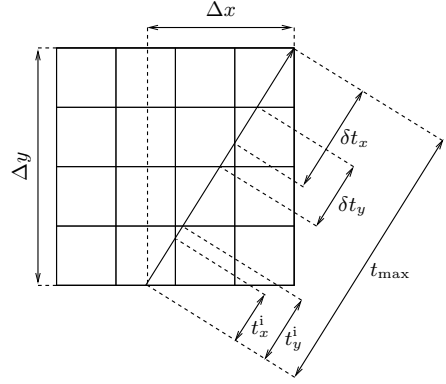}}
  \caption{
Explanation of different quantities used in the fast voxel traversal method.
Shown is a single patch with a local long characteristic ray section.
}
  \label{fig:fastVoxel}
\end{figure}
%
%
Here we briefly discuss the `fast voxel traversal algorithm' from \citet{1987Eurographics..1A}.
We have used this algorithm twice in our method, once to ray trace through the cells (`voxels') of a single patch (local long characteristics, Sect. \ref{sect:rayTraceLocal}), and once to ray trace the patch-mapping (hybrid characteristics, Sect. \ref{sect:hybridChar}).
The idea behind the algorithm is to keep track of three different ray parameters $t_x$, $t_y$, and $t_z$, one for each co-ordinate direction, and to use these to determine how to step from cell to cell through the patch, ensuring that all cells cut by the ray are visited (see Fig. \ref{fig:fastVoxel}).
First, values for the increments in ray parameter $t$ needed to step from cell to cell in the $x$-, $y$-, and $z$-direction, indicated by $\delta t_x$,  $\delta t_y$, and $\delta t_z$, respectively, are determined:
\begin{equation}
  \delta t_x=t_\mathrm{max}/\Delta x \; ,
  \delta t_y=t_\mathrm{max}/\Delta y \; ,
  \delta t_z=t_\mathrm{max}/\Delta z \; ,
\end{equation}
where $t_\mathrm{max}=\sqrt{\Delta x^2+\Delta y^2+\Delta z^2}$ is the final ray parameter (i.e. the total path length of the ray).
Next, the ray parameters $t_x$, $t_y$, and $t_z$ are set to their respective initial values, indicated by $t_x^\mathrm{i}$, $t_y^\mathrm{i}$, and $t_z^\mathrm{i}$, after which a loop is entered where the {\em minimum} of these three values is determined.
This gives the co-ordinate direction in which the cell lies that is to be visited next by the ray.
For example, if \verb|min|$(t_x,t_y,t_z)=t_x$, the next cell the ray will enter lies in the x-direction, and we have to increment the ray parameter for the x-direction accordingly, i.e. $t_x=t_x+\delta t_x$.
We loop as long as all ray parameters are smaller than the final ray parameter $t_\mathrm{max}$.
As a by-product, the algorithm produces the path length of the ray section for each cell that is crossed, which is obtained by subtracting the previous from the current ray parameter.
%
%
\section{\\Ray tracing a single patch: short characteristics}
\label{app:shortChar}
\begin{figure}
\psfrag{1}{$1$}
\psfrag{2}{$2$}
\psfrag{3}{$3$}
\psfrag{4}{$4$}
\resizebox{\hsize}{!}{\includegraphics[width=10cm]{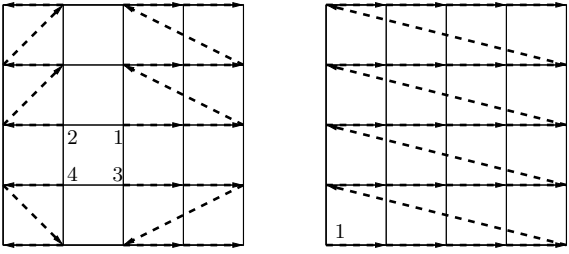}}
  \caption{
Two examples of short characteristics sweeping sequences for a single patch that may occur in practice.
For the illustration on the left, the source is located inside the patch at the starting point of curve $1$.
In this case the four space filling curves shown should be swept in the order indicated.
For the patch on the right, the source is external to the patch, and lies in the direction of the lower left corner, so there is only one curve that needs to be swept.
}
  \label{fig:sweepSeq}
\end{figure}

As an alternative to the `fast voxel traversal algorithm' for ray tracing a single patch as presented in Sect. \ref{sect:rayTraceLocal}, we here briefly describe the short characteristics method, which could be used for the same purpose.
Since the method of short characteristics uses interpolation from neighbouring cells, upwind values need to be available at all times, so cells need to be swept in a certain order.
This sweeping sequence is determined by the physical location of the patch relative to the source position (see Fig. \ref{fig:sweepSeq}).
%
%
%
%
Using the known physical location of the source, the geometrical path length of the ray section that crosses a cell is calculated for every cell contained in the patch.
The short characteristics method then sweeps the patch in a direction away from the source, interpolating upwind column density contributions for each cell along the way.

\begin{figure}
\psfrag{c}{$c$}
\psfrag{c1}{$c_1$}
\psfrag{c2}{$c_2$}
\psfrag{c3}{$c_3$}
\psfrag{c4}{$c_4$}
\psfrag{r}{$\Delta r$}
\psfrag{d}{$d$}
\psfrag{d1}{$d_1$}
\psfrag{d2}{$d_2$}
\resizebox{\hsize}{!}{\includegraphics[width=10cm]{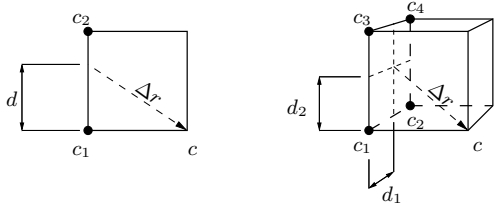}}
  \caption{
Illustration of the interpolation scheme for a single cell used in the short characteristics method for the 2D (left) and 3D (right) case.
}
  \label{fig:shortCharInt}
\end{figure}

For the two-dimensional case, Fig. \ref{fig:shortCharInt} illustrates which two cells, indicated by $c_1$ and $c_2$, are used in this interpolation.
%
%
%
%
Simple linear interpolation weights
\begin{equation}
w_1=1-d; \: w_2=d
\end{equation}
could be used to arrive at the column density contribution at cell $c$, using
\begin{equation}
\Delta N_c = \sum_i w_i \Delta N_i + \Delta r\, n\: ,
\end{equation}
with $\Delta N_i$ the upwind values of column density that need to be interpolated, $d$ the normalized distance from $c_1$ to the location where the ray pierces the line connecting $c_1$ and $c_2$, $\Delta r$ the physical path length of the short characteristic ray section, and $n$ the number density inside the cell crossed by this short characteristic.

For the three-dimensional case the ray pierces a cell face, so four instead of two quantities need to be interpolated.
The normalized weights are chosen to correspond to the partial areas of the cell face defined by the corners of this face and the location at which the ray leaves the cell (cf. Fig. \ref{fig:shortCharInt}):
\begin{equation}
\begin{array}{ll}
w_1=(1-d_1)(1-d_2); & w_2=d_1\, (1-d_2); \\
w_3=(1-d_1)\, d_2;  & w_4=d_1\, d_2 \: .
\end{array}
\end{equation}
%
%
\bibliographystyle{aa}
\bibliography{hybridCharMethodPaper}
%
%
\end{document}